\documentclass[pra,twocolumn,showpacs,amsmath,amssymb]{revtex4-2} %

\usepackage{csquotes}
\usepackage{epsfig,amsmath}
\usepackage{subfigure}
\usepackage{graphicx}
\usepackage{dcolumn}
\usepackage{stmaryrd}
\usepackage{mathrsfs}
\usepackage{pifont}
\usepackage{amsthm}
\usepackage{amssymb}
\usepackage{mathtools}
\usepackage{bm}
\usepackage{latexsym}
\usepackage[colorlinks=true,linkcolor=blue,citecolor=blue]{hyperref}
\usepackage{color}
\usepackage{hyperref}
\usepackage{xurl} 

\DeclareMathOperator*{\Motimes}{\text{\raisebox{0.25ex}{\scalebox{0.8}{$\bigotimes$}}}}

\theoremstyle{plain}

\def\pra#1{{ Phys.\ Rev. A\/} {\bf#1}}

\def\prl#1{{ Phys.\ Rev.\ Lett.} {\bf#1}}

\begin{document}

\title{One Polynomial Strategy for Computing Local Projections on Square-Lattice Cluster States: a Primary Note}

\author{Nyau Fisn}
\thanks{Contact email address: yetherphaysun@gmail.com.}

\author{Houren Fu}
\thanks{email address: saintpal@163.com}

\begin{abstract}
Quantum computing has attracted a lot of attention in recent years. It is one of the promising candidates for the next-generation computing paradigms. Basically, there are two technical lines to realize quantum computing. One is composing the unitary operators of a few qubits to achieve general unitary operators on an arbitrary number of qubits, known as the approach of quantum circuits. The other one focuses on preparing quantum cluster states and performing the computation by measuring the states with a particular basis, known as measurement-based quantum computing or one-way quantum computing. The two strategies have been proven to be equivalent to each other. This note aims to discuss the strategies for computing the local projections on square-lattice cluster states. Seemingly, one strategy for the computation could require both polynomial steps and memories. In particular, if the number of qubits in a square-lattice is denoted by $N$, the step number for computing an arbitrary local projection on the state could be proportional to a polynomial of $N$ under a bounded cost of memory. Consider that the square-lattice cluster states are one kind of universal computing resource, the results might be helpful for 
understanding the computational advantages of quantum algorithms, as well as the limits of the numerical analysis on other 
relevant quantum models. Because the results are not peer reviewed, all the criticisms, especially those on the flaws of the details, are sincerely welcomed.   
\end{abstract}


\maketitle
\section{Preface}\label{sec:Pre}
To begin with, this note would not look like a regular paper for most journals. The topic of the note might be related to a wide range of works, while the note does not have a complete reference. It would be completed if the details of the work prove to be of sufficient significance. 

The idea of quantum computing is usually thought to originate from the talk of Richard P. Feynman, which is published as a lightly edited transcript \cite{Feynman1982} later. In the talk, Feynman clearly states his goal, which is to simulate quantum systems using resources that scale well with the size of the system. The formalized notion of a quantum computer was given by David Deutsch in 1985 \cite{Deutsch1985}. It started the discussion on whether quantum computers might have an advantage over classical computers. Several remarkable progress on the issues include the case found by Umesh Vazirani and his student Ethan Bernstein (a superpolynomial speed-up \cite{Bernstein1997}), the case found by Daniel Simon (exponential speed-up \cite{Simon1997}), the discrete logarithm and integer factorization algorithm found by Peter Shor (exponential speed-up \cite{Shor1999}), and the fast search algorithm found by Lov K. Grover \cite{Grover1997}. Also, to achieve a large-scale quantum computer, error correction code is unavoidable. The first one of this code was also given by Shor \cite{Shor1995}, dating back to 1995. The short history of quantum computing has many exciting moments. One can find a very interesting review on the moments in Ref. \cite{Preskill2021}. 

Given a thought about the original description of Feynman from the current perspective of information science, one might generate different options. Yes, indeed, it is a great idea to compute quantum evolutions using a physical system that evolves under the ``quantum rules''. However, is this the optimal way to perform such a computation? Perhaps not. This can be seen from the origin of the difficulties in performing the calculations. By far, the models of the physical world are built on mathematics, and the same is true for those of quantum cases. The basic representation strategy for the states of quantum individuals is clear and well accepted, which is the elements in the space spanned by the solutions of particular equations, such as Schr{\"o}dinger equation, Klein-Gordon equation, Dirac equation, etc. Lucky for us, such a kind of space is a Hilbert space, an infinite-dimensional space with convergent inner product. Hence, its properties can be roughly sketched by a linear space with finite dimensions, easy for us to understand. On such a foundation, the ``difficulties of evaluating quantum evolutions'' might be viewed as the ``difficulties of computing the transformation of the elements in the Hilbert space''. In today's view, the computation of a function is just an action that firstly maps the function to a suitable basis of space and secondly performs the transforms on the basis to obtain the equivalent results of the function. As such, the question on the optimality could be expressed in a more specific way: {\it can a space--not necessarily a linear space--be found for framing the function of quantum evolution such that the mimic of the quantum evolution can be expressed by a small number of simple basis transformations of the space?} \cite{Footnote1} 

The authors of the note are not majoring in mathematics, so the above question might not be mathematically appropriate. However, we try to express the key point. One might think of classical computing and quantum computing as examples. In the classical computation of quantum evolution, the basis for achieving the goal is set to be the compositions of binary digits, or bits. It has already shown that such a basis is low-efficient in expressing quantum states, not to mention the mimic of the evolution by using binary transformations. In contrast, the basis in quantum computing, which closely resembles the basis of the evolution to be mimicked, should be inherently more efficient. This is because the mapping of the basis to the mimicked quantum evolution can be constructed by only unitary transformations, without any additional factors. 
This is the understanding of the quantum advance in the authors' view. It seems that the evidence for the quantum advance only shows the benefit of choosing the quantum states as the basis for the computations, not the optimality of doing so. There are potential choices of the basis that could be better for computing the outcomes of quantum evolutions. Maybe, there have already been interesting conclusions in the direction (such as the researches on graphic processing units or tensor processing units), which are missed by us. If so, we would like to be informed with great excitement. In fact, much attention has been drawn to developing quantum computing devices, which fundamentally improve the capability of controlling quantum particles. We think that the other approaches to the problem, such as finding better basis for the computation, or searching for novel transformations in larger well-defined function space that could help, etc, should also be given significant notice. 
The works such as Ref. \cite{Aaronson2004,Nest2010,Vidal2004,Vidal2007,Vidal2008,Redei2006,Nevile2017,Noh2020,Oh2021,Oh2024-1,Oh2024-2,Chia2018,Gilyen2018,Tang2018,Tang2021,Kalachev2021,Pan2022,Wetering2020,Gulbahar2019,Yang2024,Atallah2024,Martinez2025,Berezutskii2025,Orus2014,Wahl2023,Llima2024,Ma2024,Gabrielli2025,Klinkenberg2023} 
might be considered as relevant researches for this purpose or as the ones that at least could contribute to it.

The method introduced in the note might be deeply connected to the method mentioned above ----the representatives of many works in the relevant research fields. However, because we have not gone deeply into the above works, we were unable to express what the connection would be like at the first place. We only try to express the idea in the thinking lines of us. This informal note focuses on the computation of a special type of cluster states. Cluster states are a special type of quantum entangled states, which is the foundation of measurement-based quantum computing (MBQC) \cite{Raussendorf2001,Raussendorf2003}. Those states can be represented by graphs composed of vertices and edges. Each vertex represents a qubit state $|+\rangle=(|0\rangle+|1\rangle)/\sqrt{2}$. Each edge between two vertices represents a control-$Z$ (shortened as CZ) operator between the qubits indicated by the vertices. By measuring parts of the qubit states, the rest of the qubits will change to the output states of a certain task or algorithm, which are also the manipulation results of certain unitary gate circuits effectively. Hence, universal quantum computations, which are frankly described by unitary transformations on quantum states, can be realized by measuring cluster states. 

\begin{figure}[htbp]
\centering
\includegraphics[width=2.8in]{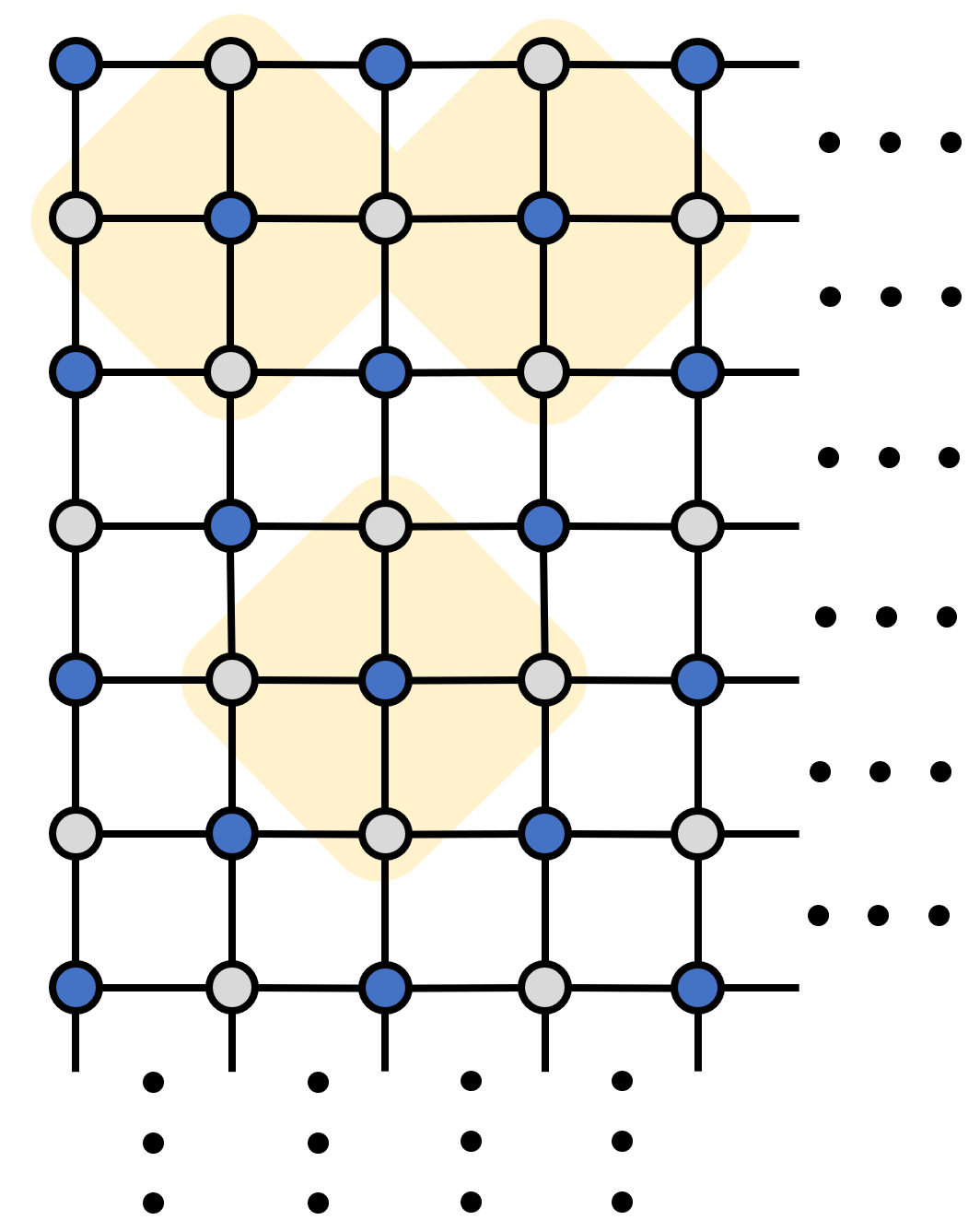}
\caption{The graph of a 2DSLCS, with the qubits as the controls (targets) being colored gray (blue). The yellow areas are examples of the basic components in the graph, which can nearly generate the whole graph 
}\label{Fig:2DCS}
\end{figure}

Specified by the topology of connections, there are infinitely many cluster states. In fact, a cluster state with a square lattice structure is already sufficient to perform universal quantum computation. The graph of a 2D square-lattice cluster state (2DSLCS) is shown in Fig. \ref{Fig:2DCS}. Obviously, each qubit, except the one in the corner or the boundary, has four qubits directly connected to it. The fact indicates that the state of each qubit is at most controlled by four qubits, and each qubit controls at most four qubits at the same time. This is also a fundamental property of a square lattice. As shown in Fig. \ref{Fig:2DCS}, for a 2DSLCS, one can always divide the vertices of the lattice into two categories. One includes the qubits severing as the controls, the other includes the qubits serving as the targets. In Fig. \ref{Fig:2DCS}, the control qubits are colored gray, and the target qubits are colored blue. Due to the property of a CZ gate, the control and the target can be exchanged. Therefore, the colors of the blue and gray vertices in Fig. \ref{Fig:2DCS} can also be exchanged. This is like the concept of the A-B sub-lattice in condensed matter physics.

Particularly, if one denotes the set of connected qubits by $G$, a 2DSLCS can be given by
\begin{equation}\label{eq:ProdForm}
    \prod_{(m,n)\in G}CZ_{m,n} \left(\Motimes_{p=1}^N |+\rangle_p\right).
\end{equation}
$N$ denotes the total number of qubits. $CZ_{m,n}$ denotes the control-$Z$ gate acting on the $m$th and $n$th qubit. $(m,n)$ denotes a pair of qubit indices involved in a control-$Z$ gate (also an edge of the graph of a 2DSLCS). Following the above conclusion, one can categorize the qubits in Eq. (\ref{eq:ProdForm}) into two types, the control and the target qubits. Suppose that the number of control qubits is $K$. Then, the number of target qubits is $N-K$. Because there are no direct connections among the control qubits, the total state of the control qubits is a tensor product state running over all the binary bases. Furthermore, a 2DSLCS can be reformed by using the above categorizations, 
\begin{equation}\label{eq:2form}
    \begin{split}
        \frac{1}{2^{K/2}}\sum_{j_1,j_2,\ldots,j_K=0}^1&|j_1j_2\ldots j_K\rangle\Motimes_{q=1}^{N-K}[(1-\alpha_q)|+\rangle\\
        &+\alpha_q|-\rangle].
    \end{split}
\end{equation}
$j_1,j_2,\dots,j_K$ are binary integers, being $0$ or $1$. For one set of binaries $j_1,\ldots,j_K$, a unique set of $a_1,\ldots,a_{N-K}$ can be given, certifying that the related target qubits are $|+\rangle$s or $|-\rangle$s. Using the form of $(1-\alpha_q)|+\rangle+\alpha_q|-\rangle$ with integer $q$, one can find $\alpha_q$ satisfies $\alpha_q=(j_{q_1}+j_{q_2}+j_{q_3}+j_{q_4})\pmod{2}$. $j_{q_1}$, $j_{q_2}$, $j_{q_3}$, and $j_{q_4}$ are the index of the control qubits for the $q$th target qubit. Frankly, the expression of $a_q$ actually counts the number of control qubits in $|1\rangle$. A simple example would be a five-qubit cluster state, with one qubit connecting to the rest individual four qubits. It is expressed by 
\begin{equation}\label{eq:5qeg}
    \frac{1}{2^4}(|0000\rangle\otimes|+\rangle+|0001\rangle\otimes|-\rangle+\cdots+|1111\rangle\otimes|+\rangle).
\end{equation}
For one basis state of Eq. (\ref{eq:5qeg}), its fifth qubit would be $|+\rangle$ if and only if an even number of qubits among the first four is $|0\rangle$. Otherwise, the fifth qubit would be $|-\rangle$. More details of the state are given in Section \ref{subsec:CSCS}.

Of course, if one cares more about the computational tasks encoded by Eq. (\ref{eq:2form}), the projections of Eq. (\ref{eq:2form}) should be more important. In MBQC, the algorithm for a given task is performed by setting the projection basis of the state like Eq. (\ref{eq:2form}). The connections of the state can also be modified by suitable projections. Suppose that the projector is denoted by $\Motimes_{p=1}^{N}\left(C_p\langle0|+S_p\langle1|\right)$, where $C_p=\cos\theta_p$ and $S_p=e^{i\varphi_p}\sin\theta_p$. Then, the projection is given by,
\begin{equation}\label{eq:proj}
\begin{split}
&\left[\Motimes_{p=1}^{N}\left(C_p\langle0|+S_p\langle 1|\right)\right] \left[\prod_{(m,n)\in G}CZ_{m,n} \left(\Motimes_{p=1}^N |+\rangle_p\right)\right]     \\
=&\sum_{j_1,j_2,\ldots,j_K=0}^1 \{\prod_{s=1}^K [(1-j_s)C_s+j_sS_s]\prod_{q=1}^{N-K}[(1-\alpha_q)   \\
& \times(C_q+S_q)+\alpha_q(C_q-S_q)]\}.
\end{split}
\end{equation}
In quantum computation, to evaluate Eq. (\ref{eq:proj}) is not easy because, in general, there are many terms in the equation (the order is about $2^N$). The following note focuses on the computation of Eq. (\ref{eq:proj}), which is an output of the MBQC based on 2DLSCS. The main idea is inspired by the binomial theorem, which can be expressed by
\begin{equation}\label{eq:BinomialT}
   \sum_{j_1,j_2,\ldots,j_K=0}^1\prod_{s=1}^K [(1-j_s)C_s+j_sS_s]=\prod_{s=1}^K (C_s+S_s).
\end{equation}
It can be noticed that the computing consumption of the two sides of Eq. (\ref{eq:BinomialT}) are different. On the left-hand side, there are $2^K$ terms in total. Hence, one needs to perform $2^K-1$ additions and $(2^K-1)(K-1)$ multiplications to evaluate the expression. However, one only needs to perform $K-1$ additions and $K-1$ multiplications to evaluate the expression on the right-hand side. It should have been known for long that factorizing a polynomial provides a good alternative to compute it. In the following sections, we go into specific examples by using the above idea. 

\section{Factorizing the Projections of Several Examples}\label{sec:CS}

In spite of strict logic lines based on theorems, we provide only examples. Such an arrangement might make it easier for us to express the idea, not caring much about setting the organization of the note. However, we think that the examples considered here are either representative, or easy to be extended, or share both merits, so that they can be applied to a very broad range of cases.

\subsection{Line-Shape Cluster States}

\begin{figure}[htbp]
\centering
\includegraphics[width=3.1in]{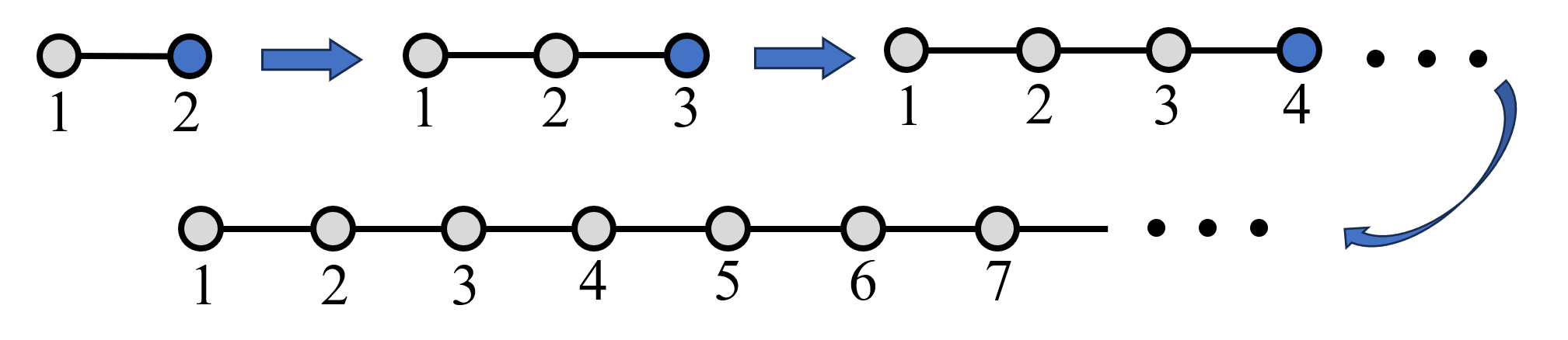}
\caption{The extension of line-shape cluster states.}\label{Fig:LSCS}
\end{figure}

Line-shape cluster states are one type of simple cluster states. Their graphs are shown in Fig. \ref{Fig:LSCS}. 

For the simplest one of such a simple case---the initial of the chain, it can be expressed by
\begin{equation}
    CZ_{1,2}|+\rangle_1|+\rangle_2=\frac{1}{\sqrt{2}}(|0\rangle_1|+\rangle_2+|1\rangle_1|-\rangle_2).
\end{equation}
Such a state is also known as one of the Bell states. Using the above projectors, one has
\begin{equation}\label{eq:FactorLSCS-2q}
    \begin{split}
        &\left[(C_1\langle0|+S_1\langle1|)(C_2\langle0|+S_2\langle1|)\right]CZ_{1,2}|+\rangle_1|+\rangle_2\\
        =&\frac{1}{2}[C_1(C_2+S_2)+S_1(C_2-S_2)].
    \end{split}
\end{equation}
Obviously, the polynomial $C_1(C_2+S_2)+S_1(C_2-S_2)$ cannot be written in a product form like $(aC_1+bS_1)(cC_2+dS_2)$, for $a,b,c,d\in\mathbb{R}$ \cite{Footnote-1}. However, we might consider matrix $a,b,c$, and $d$. If the 2-by-2 identity matrix is denoted by $I$, and the Pauli-Z matrix is denoted by $Z$, then one has,
\begin{equation}
\begin{split}
    &C_1(C_2+S_2)+S_1(C_2-S_2)\\
    =&Tr\{(UC_1+DS_1)(IC_2+ZS_2)\},
\end{split}
\end{equation}
where $U=(I+Z)/2$ and $D=(I-Z)/2$ (in the language of qubits, $U$ and $D$ denotes matrix representation of $|0\rangle\langle0|$ and $|1\rangle\langle1|$). $Tr\{\cdot\}$ means taking the trace. Such a trick also draws inspiration from the investigations on mimicking quantum entanglements with classical physics \cite{Sun2015,Sun2022}. It might also have been involved or discussed in the researches on polynomial factorization (such as those in Galois theory). The author of the note would not dive into it and concentrate on the current topic. In addition, we would like to mention that the trace of multiple diagonal matrices could be considered as a generalized {\it dot product} for vectors. The dot product for vectors is one kind of mapping that maps two vectors to a real number. The generalized dot product mentioned here is a more complicated mapping that maps three or more vectors to a real number. Also, because diagonal matrices commute with each other, the evaluation of their multiplication can be performed in a suitable order, providing potential benefit for the computation. 

\begin{figure}
\centering
\includegraphics[width=3.4in]{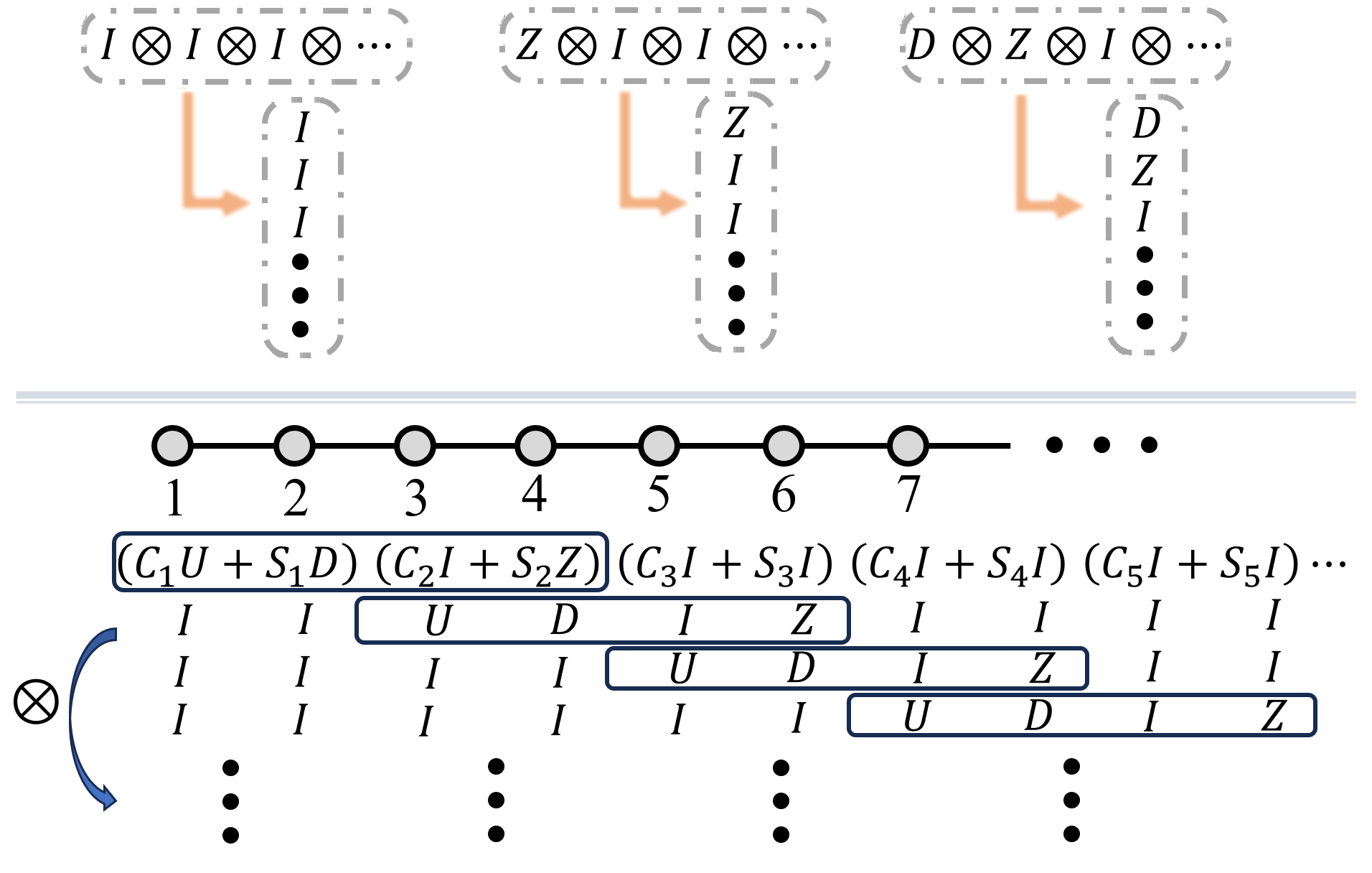}
\caption{An illustration of the factorized projection of a line-shape cluster state. The upper panel shows the graph representation of the tensor products. The lower panel shows the factorized polynomial using the representation. The black boxes in the lower panel mark the matrices that needs to be specially noticed in the multiplication.}\label{Fig:FactorLSCS}
\end{figure}

Going further, one can extend the above strategy. For the state $ CZ_{1,2}CZ_{2,3}|+\rangle_1|+\rangle_2|+\rangle_3$ (the second state of the chain in Fig. \ref{Fig:LSCS}), the projection is given by 
\begin{equation}\label{eq:FactorLSCS-3q}
    \begin{split}
        &\left[\Motimes_{p=1}^{3}\left(C_p\langle0|+S_p\langle 1|\right)\right]CZ_{1,2}CZ_{2,3}|+\rangle_1|+\rangle_2|+\rangle_3\\
        =&\frac{1}{2^{3/2}}(C_1[C_2(C_3+S_3)+S_2(C_3-S_3)]\\
        &~~~~~~~+S_1[C_2(C_3+S_3)-S_2(C_3-S_3)]).
    \end{split}
\end{equation}
The factorization of the projection result can be given by 
\begin{equation}
\begin{split}
    &\frac{1}{2^{3/2}}Tr\{[(U\otimes I)C_1+(D\otimes I)S_1][(I\otimes U)C_2\\
    &~~~~~~~~~+(Z\otimes D)S_2][(I\otimes I)C_3+(I\otimes Z)S_3]\}.
\end{split}
\end{equation}
It is easy to check the validity of the above factorization. Along the line, for a line-shape cluster state composed of $N$ qubits, the factorization of the projection can be given by
\begin{equation}\label{eq:FactorLSCS}
\begin{split}
    &\frac{1}{2^{N/2}}Tr\{[(U\otimes I^{\otimes (N-2)})C_1+(D\otimes I^{\otimes (N-2)})S_1]\\
    &\cdot[(I\otimes U\otimes I^{\otimes (N-3)})C_2+
    (Z\otimes D\otimes I^{\otimes (N-3)})S_2]\\
    &\cdot[(I\otimes I\otimes U\otimes I^{\otimes (N-4)})C_3+(I\otimes Z\otimes D
    \otimes I^{\otimes (N-4)})\\
    &\cdot S_3]\cdots[(I^{\otimes (N-2)}\otimes I)C_N+(I^{\otimes (N-2)}\otimes Z)S_N]\}.
\end{split}
\end{equation}
Also, it is relatively easy to check the factorized form. To visualize the form a little bit, one might consider the graph in Fig. \ref{Fig:FactorLSCS}, which arranges the tensor product in a column and omits the ``$\otimes$'' notations. A benefit of the graph is that it helps to align the matrix that should be especially noticed in the multiplication. Besides, one might think Eq. (\ref{eq:FactorLSCS}) is hard to evaluate at the first glance. Actually, it can be evaluated in a recursive way. Starting from the first two terms of Eq. (\ref{eq:FactorLSCS}), if the irrelevant factors (such as the total constant $2^{-N/2}$) are omitted (the same as below), one has
\begin{equation}\label{eq:LSCS-C1}
\begin{split}
    Tr\{&[(U\otimes I)C_1+(D\otimes I)S_1]\\
    &~~~~\cdot[(I\otimes U)C_2+(Z\otimes D)S_2]\}\\
    =&Tr\{[(C_1C_2U+S_1C_2D)\otimes U\\
    &~~~~~~+(C_1S_2U-S_1S_2D)\otimes D]\}\\
    =&Tr\{P_1\}Tr\{U\}+Tr\{Q_1\}Tr\{D\}.
\end{split}
\end{equation}
where $P_1=C_1C_2U+S_1C_2D$ and $Q_1=C_1S_2U-S_1S_2D$ (also, notice that $UZ=U$ and $DZ=-D$). Next, consider the first three terms of Eq. (\ref{eq:FactorLSCS}) and also omit the irrelevant factors, one has, 
\begin{equation}\label{eq:LSCS-C2}
\begin{split}
    Tr\{&[(U\otimes I\otimes I)C_1+(D\otimes I\otimes I)S_1]\\
    &~~~~\cdot[(I\otimes U\otimes I)C_2+(Z\otimes D\otimes I)S_2]\\
    &~~~~\cdot[(I\otimes I\otimes U)C_3+(I\otimes Z\otimes D)S_3]\}\\
    =&Tr\{[P_1\otimes U\otimes I+Q_1\otimes D\otimes I]\\
    &~~~~\cdot[(I\otimes I\otimes U)C_3+(I\otimes Z\otimes D)S_3]\}\\
    =&Tr\{C_3(P_1\otimes U+Q_1\otimes D)\}Tr\{U\}\\
    &~~~+Tr\{S_3(P_1\otimes U-Q_1\otimes D)\}Tr\{U\}\\
    =&Tr\{P_2\}Tr\{U\}+Tr\{Q_2\}Tr\{D\},
\end{split}
\end{equation}
where $P_2=C_3(P_1\otimes U+Q_1\otimes D)$ and $Q_1=S_3(P_1\otimes U-Q_1\otimes D)$. Notice that $Tr\{P_2\}=C_3(Tr\{P_1\}+Tr\{Q_1\})$ and $Tr\{Q_2\}=S_3(Tr\{P_1\}-Tr\{Q_1\})$, indicating that $Tr\{P_2\}$ and $Tr\{Q_2\}$ can be evaluated based on $Tr\{P_1\}$ and $Tr\{Q_1\}$. Using the strategy again and again, the trace in Eq. (\ref{eq:FactorLSCS}) can be expressed by 
\begin{equation}\label{eq:LSCS-C3}
\begin{split}
    &Tr\{[P_{N-2}\otimes U+Q_{N-2}\otimes D][(I^{\otimes (N-2)}\otimes I)C_N\\
    &~~~~~~~+(I^{\otimes (N-2)}\otimes Z)S_N]\}\\
    =&Tr\{(C_N+S_N)P_{N-2}\otimes U+(C_N-S_N)\\
    &~~~~~~~\cdot Q_{N-2}\otimes D\}\\
    =&(C_N+S_N)Tr\{P_{N-2}\}+(C_N-S_N)Tr\{Q_{N-2}\}.
\end{split}
\end{equation}
According to the recursive relations, one has
\begin{equation}\label{eq:LSCS-C4}
\begin{split}
    &Tr\{P_{N-2}\}=C_{N-1}(Tr\{P_{N-3}\}+Tr\{Q_{N-3}\}),\\
    &Tr\{Q_{N-2}\}=S_{N-1}(Tr\{P_{N-3}\}-Tr\{Q_{N-3}\}).
\end{split}
\end{equation}
It might be noticed that, by using Eq. (\ref{eq:LSCS-C3}) and Eq. (\ref{eq:LSCS-C4}), the polynomial of the projection (\ref{eq:FactorLSCS}) can be evaluated by performing $O(N)$ times additions and multiplications. This is an efficient computation, but is not a surprising result. Actually, the line-shape cluster state is equivalent to the rotations of a single qubit, which can be simulated classically. Another interesting thing is that the matrix coefficients and the factorization seem not to be necessary for obtaining the recursive relations. Simply observing the symmetry of Eq. (\ref{eq:FactorLSCS-2q}), (\ref{eq:FactorLSCS-3q}), and (\ref{eq:FactorLSCS}), one might find the clues about the recursive relations directly. While, in the next examples, it might be seen that the matrix formulations and the factorization are helpful for finding efficient computations of more complicated cases. 

\subsection{Cross-shape Cluster States}\label{subsec:CSCS}
\begin{figure}
\centering
\includegraphics[width=3.4in]{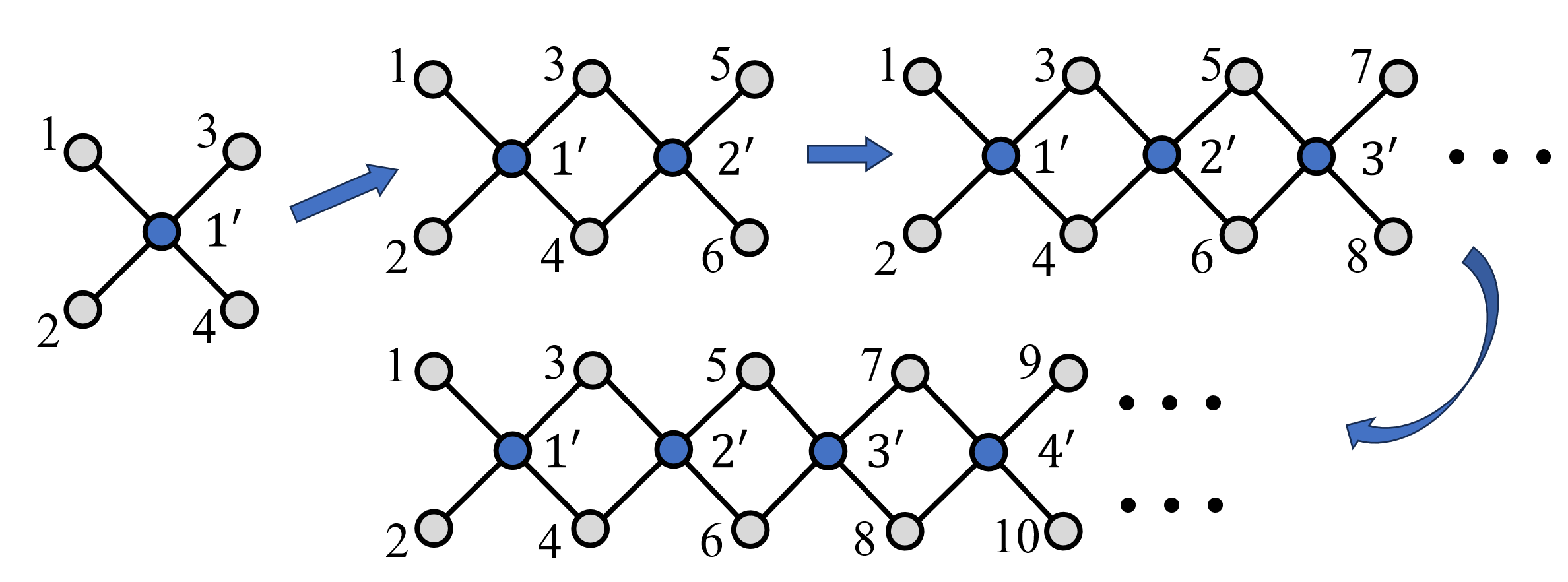}
\caption{A 1-D extension of the cross-shape state.}\label{Fig:CSCS}
\end{figure}

Cross-shape cluster states are also one type of simple cluster states. Their graphs are shown in Fig. \ref{Fig:CSCS}. There are many ways to extend the state to a larger scale. The method given in Fig. \ref{Fig:CSCS} is a one-dimensional extension, which could be helpful for understanding the case.  

\begin{figure}
\centering
\includegraphics[width=3.4in]{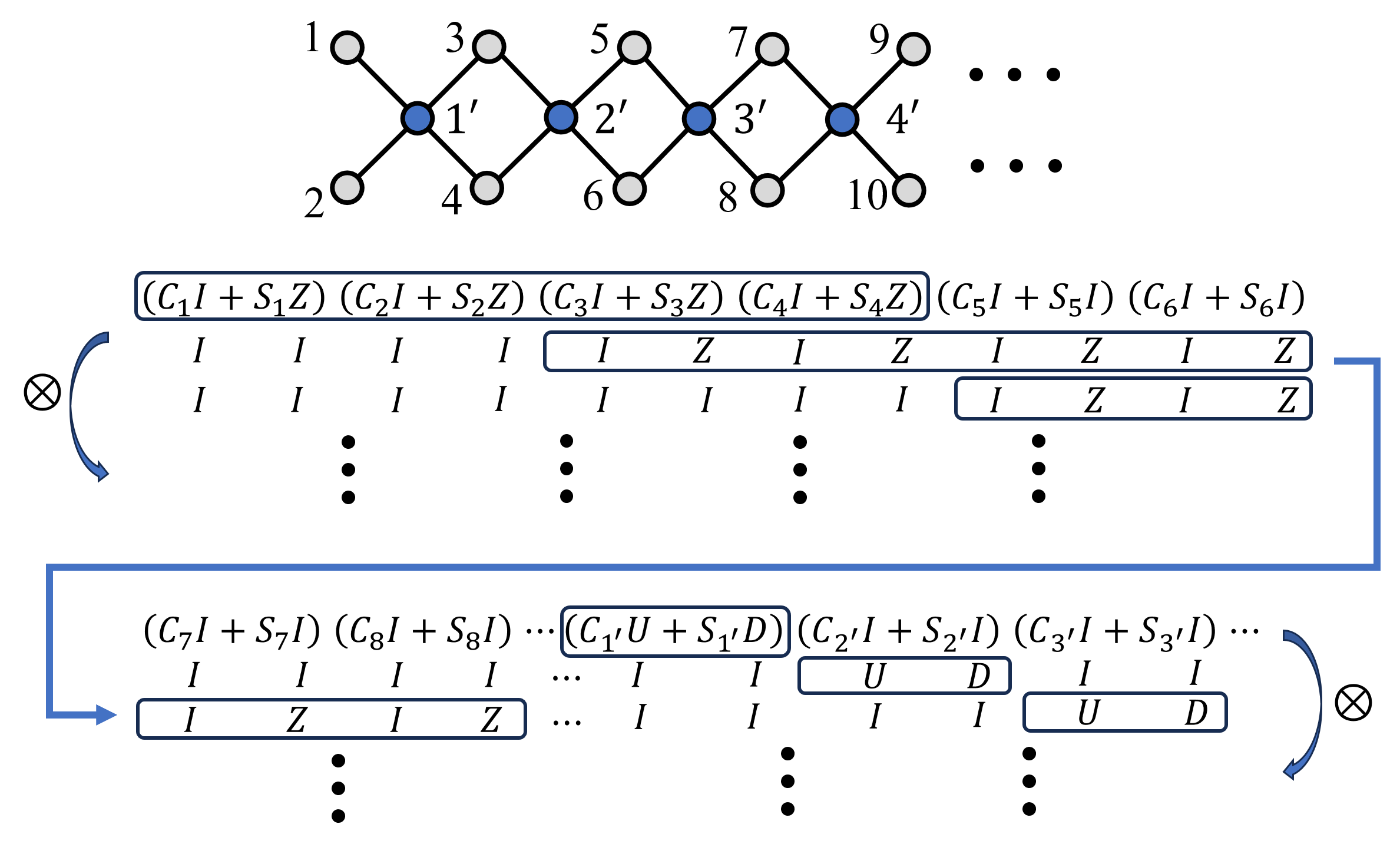}
\caption{A illustration of the factorized projection of a cross-shape-shape cluster state using the same representation of Fig. \ref{Fig:FactorLSCS}. The black boxes also mark the matrices that needs to be specially noticed in the multiplication. The arrow marks the places that are connected.}\label{Fig:FactorCSCS}
\end{figure}

The simplest one in such a case is also given in the initial of the chain, expressed by 
\begin{equation}\label{eq:GHZ}
\begin{split}
    &CZ_{1',1}CZ_{1',2}CZ_{1',3}CZ_{1',4}|+\rangle_{1'}|+\rangle_1\\
    &\otimes|+\rangle_2|+\rangle_3|+\rangle_4\\
    =&\frac{1}{\sqrt{2}}(|0\rangle_{1'}|+\rangle_1|+\rangle_2|+\rangle_3|+\rangle_4\\
    &+|1\rangle_{1'}|-\rangle_1|-\rangle_2|-\rangle_3|-\rangle_4).
\end{split}
\end{equation}
The use of $1'$ is to distinguish the roles of the qubits (the same as below). Actually, state (\ref{eq:GHZ}) is a five-qubit Greenberger-Horne-Zeilinger (GHZ) state. Following the notations in the previous sections, the projection of state (\ref{eq:GHZ}) can be given by
\begin{equation}\label{eq:ProjGHZ}
    \begin{split}
        &\left(C_{1'}\langle0|+S_{1'}\langle 1|\right)\left[\Motimes_{p=1}^{4}\left(C_p\langle0|+S_p\langle 1|\right)\right]\\
        &\cdot CZ_{1',1}CZ_{1',2}CZ_{1',3}CZ_{1',4}|+\rangle_{1'}|+\rangle_1\\
    &\otimes|+\rangle_2|+\rangle_3|+\rangle_4\\
        =&\frac{1}{2^{5/2}}[C_{1'}(C_1+S_1)(C_2+S_2)(C_3+S_3)(C_4+S_4)\\
        &~~+S_{1'}(C_1-S_1)(C_2-S_2)(C_3-S_3)(C_4-S_4)].
    \end{split}
\end{equation}
Using $\{I,Z,U,D\}$, the polynomial in Eq. (\ref{eq:ProjGHZ}) can also be factorized by
\begin{equation}\label{eq:FacProjGHZ}
    \begin{split}
        &Tr\{(C_1I+S_1Z)(C_2I+S_2Z)(C_3I+S_3Z)\\
        &(C_4I+S_4Z)(C_{1'}U+S_{1'}D)\}.
    \end{split}
\end{equation}
The procedures in the following are similar to those for the line-shape cluster states. For the second state given by Fig. \ref{Fig:CSCS}, a double cross-shape state, the projection is expressed by
\begin{equation}\label{eq:ProjDS}
    \begin{split}
        &\frac{1}{2^4}[C_{1'}C_{2'}(C_1+S_1)(C_2+S_2)(C_3+S_3)(C_4+S_4)\\
        &\cdot(C_5+S_5)(C_6+S_6)+S_{1'}C_{2'}(C_1-S_1)(C_2-S_2)\\
        &\cdot(C_3-S_3)(C_4-S_4)(C_5+S_5)(C_6+S_6)+C_{1'}S_{2'}\\
        &\cdot(C_1+S_1)(C_2+S_2)(C_3-S_3)(C_4-S_4)(C_5-S_5)\\
        &\cdot(C_6-S_6)+S_{1'}S_{2'}(C_1-S_1)(C_2-S_2)(C_3+S_3)\\
        &\cdot(C_4+S_4)(C_5-S_5)(C_6-S_6)].
    \end{split}
\end{equation}
It can be factorized as
\begin{equation}
    \begin{split}
        &Tr\{[C_1(I\otimes I)+S_1(Z\otimes I)][C_2(I\otimes I)+S_2(Z\otimes I)]\\
        &\cdot[C_3(I\otimes I)+S_3(Z\otimes Z)][C_4(I\otimes I)+S_4(Z\otimes Z)]\\
        &\cdot[C_5(I\otimes I)+S_5(I\otimes Z)][C_6(I\otimes I)+S_6(I\otimes Z)]\\
        &\cdot[C_{1'}(U\otimes I)+S_{1'}(D\otimes I)][C_{2'}(I\otimes U)+S_{2'}(I\otimes D)]\}.
    \end{split}
\end{equation}
Also, it is easy to check the factorization. Furthermore, for the $N$-cross-shape cluster state, the factorized form can be given by Fig. \ref{Fig:FactorCSCS}. The phrase``$N$-cross-shape cluster state'' means that the subscripts with a prime take from $1'$ to $N'$. Thus, such a state includes $3N+2$ qubits in total.

One might notice that the rule for giving the factorization is using Eq. (\ref{eq:FacProjGHZ}) for each ``cross'' and then guaranteeing that the matrix coefficients of the shared qubits are in one tensor product. The computation of the polynomial in Fig. \ref{Fig:FactorCSCS} is also similar. Taking the first four factors of the factorization as an example, one has
\begin{equation}
    \begin{split}
    &[C_1(I\otimes I^{\otimes(N-1)})+S_1(Z\otimes I^{\otimes(N-1)})][C_2(I\\
    &~~~~\otimes I^{\otimes(N-1)})+S_2(Z\otimes I^{\otimes(N-1)})]\\
    =&(C_1C_2+S_1S_2)(I\otimes I^{\otimes(N-1)})+(C_1S_2+S_1C_2)\\
    &~~~~\cdot(Z\otimes I^{\otimes(N-1)})\\
    =&\tilde{C}_1(I\otimes I^{\otimes(N-1)})+\tilde{S}_1(Z\otimes I^{\otimes(N-1)}),\\
    \end{split}
\end{equation}
and
\begin{equation}
    \begin{split}
    &[C_3(I^{\otimes2}\otimes I^{\otimes(N-1)})+S_3(Z^{\otimes2}\otimes I^{\otimes(N-1)})]\\
    &~~~~[C_4(I^{\otimes2}\otimes I^{\otimes(N-1)})+S_4(Z^{\otimes2}\otimes I^{\otimes(N-1)})]\\
    =&(C_3C_4+S_3S_4)(I^{\otimes2}\otimes I^{\otimes(N-1)})+(C_3S_4\\
    &~~~~+S_3C_4)(Z^{\otimes2}\otimes I^{\otimes(N-1)})\\
    =&\tilde{C}_2(I^{\otimes2}\otimes I^{\otimes(N-1)})+\tilde{S}_2(Z^{\otimes2}\otimes I^{\otimes(N-1)}).\\
    \end{split}
\end{equation}
By multiplying the factor $(C_{1'}U+S_{1'}D)\otimes I^{\otimes (N-1)}$ and temporarily omitting several factors, two resultant factors can be obtained. The first one is
\begin{equation}
    \begin{split}
        &[\tilde{C}_1(I\otimes I)+\tilde{S}_1(Z\otimes I)][(C_{1'}U+S_{1'}D)\otimes I]\\
        =&\tilde{C}_1(C_{1'}U+S_{1'}D)\otimes I+\tilde{S}_1(C_{1'}U-S_{1'}D)\otimes I\\
        =&\tilde{C}_1T^{+}_1\otimes I+\tilde{S}_1T^{-}_1\otimes I,
    \end{split}
\end{equation}
where $T^{+}_1=C_{1'}U+S_{1'}D$ and $T^{-}_1=C_{1'}U-S_{1'}D$. Notice that $ZT^{+}_1=T^{-}_1$. The second one is
\begin{equation}
    \tilde{C}_2(I\otimes I)+\tilde{S}_2(Z\otimes Z).
\end{equation}
These two factors go back to the form of Eq. (\ref{eq:LSCS-C1}). Multiply them, one has
\begin{equation}
    \begin{split}
        &(\tilde{C}_1T^{+}_1\otimes I+\tilde{S}_1T^{-}_1\otimes I)[\tilde{C}_2(I\otimes I)+\tilde{S}_2(Z\otimes Z)]\\
        =&(\tilde{C}_1T^{+}_1+\tilde{S}_1T^{-}_1)\otimes \tilde{C}_2I+(\tilde{S}_1T^{+}_1+\tilde{C}_1T^{-}_1)\otimes \tilde{S}_2Z\\
        =&\tilde{P}_1\otimes \tilde{C}_2I+\tilde{Q}_1\otimes \tilde{S}_2Z.
    \end{split}
\end{equation}
The rest of the factors in the factorization of Fig. \ref{Fig:FactorCSCS} can also be transformed into the corresponding factors in Fig. \ref{Fig:FactorLSCS}, by multiplying the factors in the black boxes pairwise. This transformation can be performed in $O(N)$ steps. Then, using the trick in the above case, one can also obtain the recursive relations, roughly given by
\begin{equation}\label{eq:CSCS-C1}
\begin{split}
    &Tr\{\tilde{P}_{N-2}\}=\tilde{C}_{N-2}Tr\{\tilde{P}_{N-3}\}Tr\{T_{N-2}^+\}\\
    &~~~~~~~~~~+\tilde{S}_{N-2}Tr\{\tilde{Q}_{N-3}\}Tr\{T_{N-2}^-\},\\
    &Tr\{\tilde{Q}_{N-2}\}=\tilde{S}_{N-2}Tr\{\tilde{Q}_{N-3}\}Tr\{T_{N-2}^+\}\\
    &~~~~~~~~~~+\tilde{C}_{N-2}Tr\{\tilde{P}_{N-3}\}Tr\{T_{N-2}^-\},
\end{split}
\end{equation}
where $Tr\{T_{N-2}^\pm\}=C_{(N-2)'}\pm S_{(N-2)'}$ (The head and tail of the recursion chain is not involved. One can easily obtain the whole result like in the case of line-shape cluster states). Therefore, the projection computation of the states in Fig. \ref{Fig:CSCS} can also be performed in $O(N)$ steps. 

Again, this is not a surprising result. Although the cross-shape cluster states in Fig. \ref{Fig:CSCS} look like a 2-D cluster state, they are fundamentally equivalent to the line-shape cluster states in Fig. \ref{Fig:LSCS}. This is because the extension of the state is in one direction. Meanwhile, the recursive relations can also be found by observing Eq. (\ref{eq:ProjGHZ}), Eq. (\ref{eq:ProjDS}), and the projection of $N$-cross-shape cluster states (not explicitly given here). Hence, the introduction of matrix formulation and factorization is also not necessary. However, directly finding the recursive relations in this case is much harder than in the previous case \cite{Footnote-Insert}. 

Returning to the factorization, one might find common characters shared by the above two cases: the effective multiplications of the matrices in a row of Fig. \ref{Fig:LSCS} and Fig. \ref{Fig:CSCS} are limited and independent of the total number of qubits, and most of the matrices in the tensor products are identical matrices especially when the qubit number is large.  
In the example of Fig. \ref{Fig:LSCS}, the effective multiplications of the matrices in one row involve only two bracketed factors. 
Similarly, in the example of Fig. \ref{Fig:CSCS}, the effective multiplications of the matrices in one row involve only four bracketed factors. 
We think these are the conditions for obtaining recursive relations in the computing of factorization. They lead to the result that the number of steps in the computation is proportional to the number of rows, i.e., roughly the number of qubits. Unfortunately, this is a conjecture at this point. Maybe, it is not hard to prove this in the case of line-shape cluster states and cross-shape cluster states. However, it is not obvious in the 2-D case, which is of the main interests.
\begin{figure}
\centering
\includegraphics[width=3.4in]{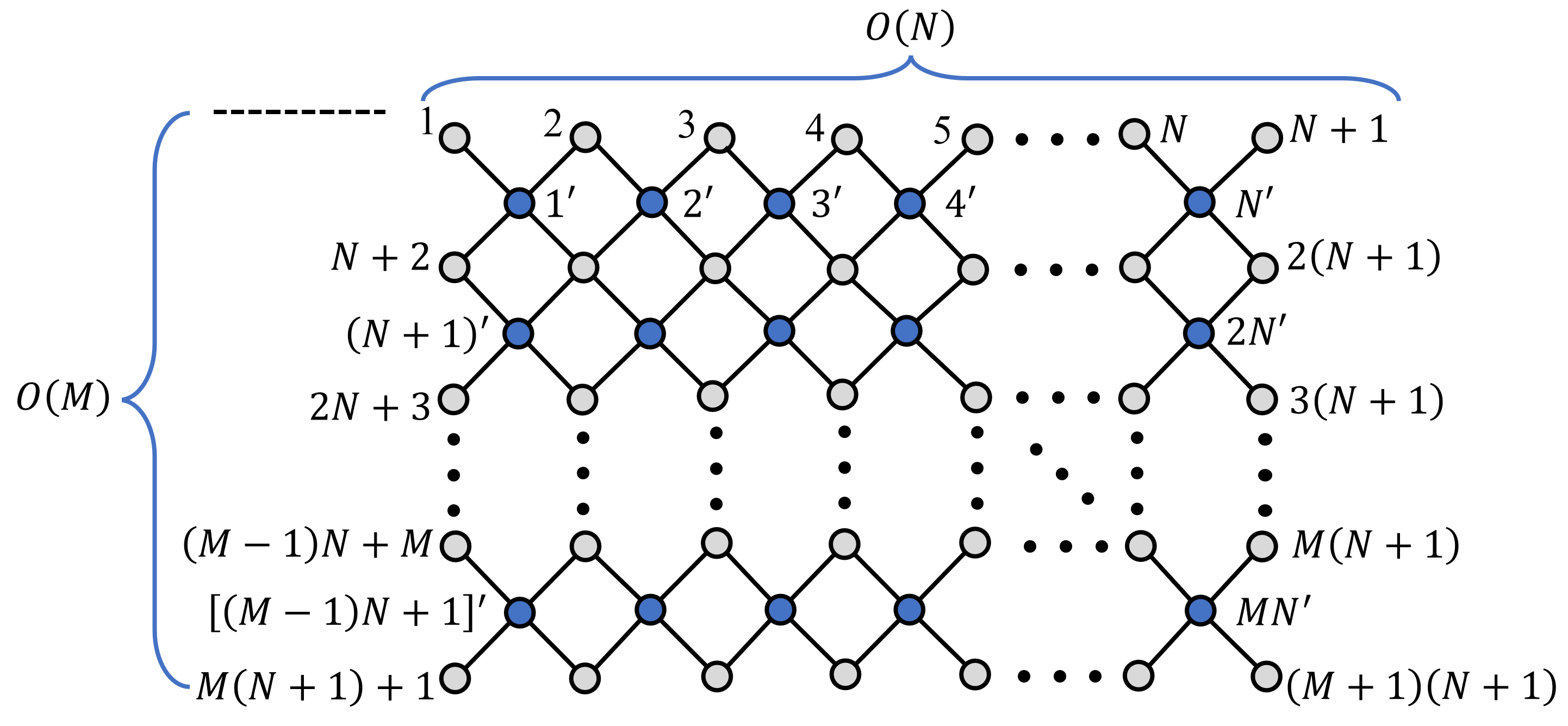}
\caption{A graphic illustration of a 2-D extension of the cross-shape cluster states. It is also a 2DSLCS as shown in Fig. \ref{Fig:2DCS}}\label{Fig:2DLSExten}
\end{figure}
\begin{figure*}
\centering
\includegraphics[width=6.7in]{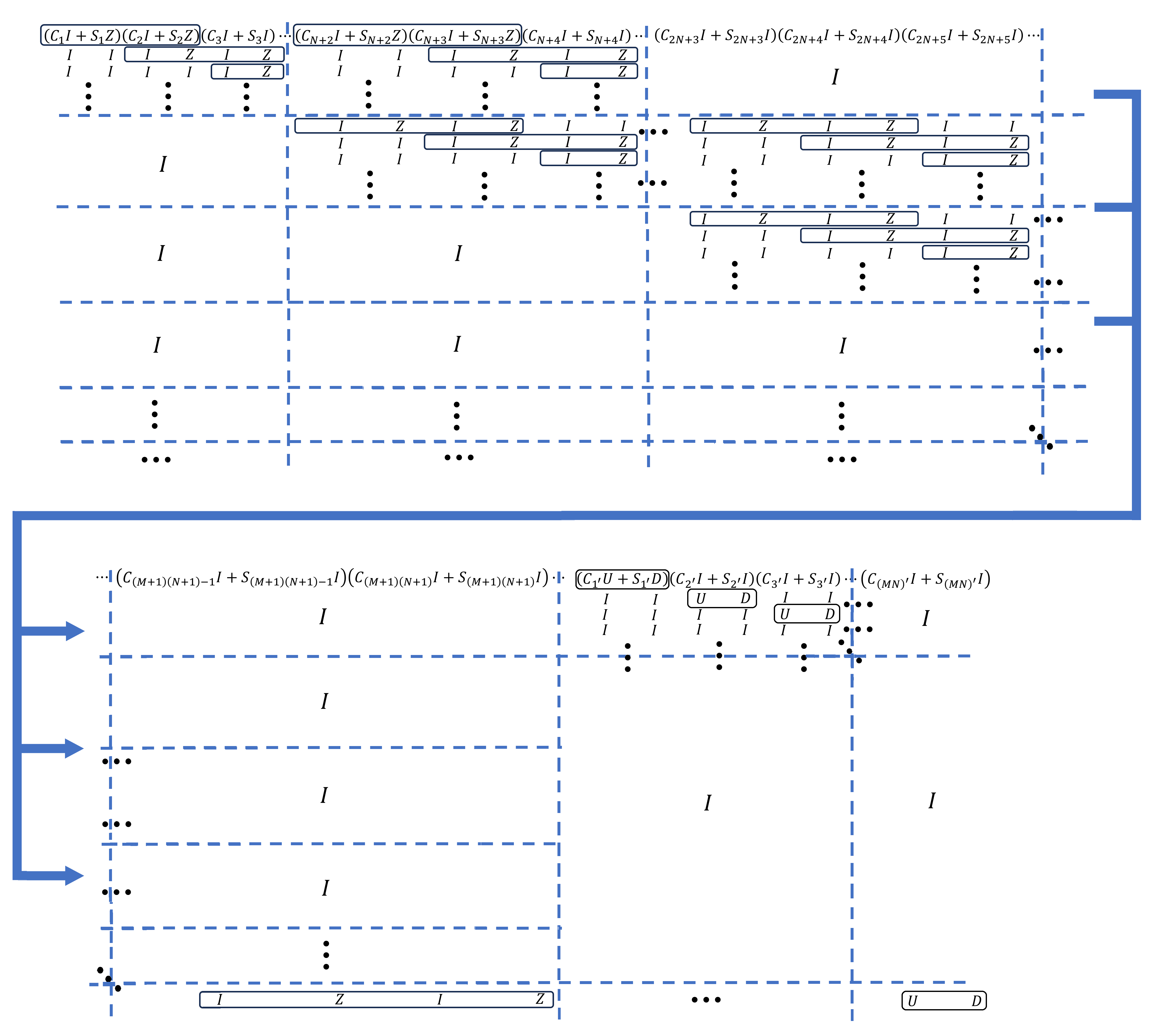}
\caption{The factorized polynomial form of  projection on the 2DSLCS in Fig. \ref{Fig:2DLSExten}. The dashed blue lines indicate the boundaries of the blocks. The repetitive elements in blocks are omitted. The black boxes also mark the matrices that needs to be specially noticed in the multiplication. The arrow marks the places that stick together.}\label{Fig:2DLSfac}
\end{figure*}

\subsection{2-D Extension of Cross-shape Cluster States}\label{subsec:2DCSCS}
As mentioned previously, the extension in Fig. \ref{Fig:CSCS} is one-dimensional. A 2-D extension of cross-shape cluster states is shown in Fig. \ref{Fig:2DLSExten}. There are $MN$ ``crosses'' in the graph of the state, so that there are $(M+1)(N+1)+MN$ qubits in total. Here, the total number of rows of qubits is $M+1$, and the total number of columns of qubits is $N+1$. Note that $N$ is used to describe the number of columns only in this subsection. In other parts of the note, $N$ is used to denote the total number of qubits.  

Using the above method to establish the factorized projection, one can obtain the polynomial given by Fig. \ref{Fig:2DLSfac}. The subscripts of $C$ and $S$ in the brackets are set exactly according to the subscripts of the qubits in Fig. \ref{Fig:2DLSExten}. The basic form of the factorization is similar to that in the previous case. However, as one might have seen, it is a bit hard to illustrate the factorization. In Fig. \ref{Fig:2DLSfac}, we divide the graph of the tensor products into blocks and omit a large number of repetitive notations in one block if it has only identity $I$s. From the blocks, one can see that the matrices that needs to be computed in the multiplication are ``diagonally'' distributed. This leads to similar characters as the factorization shown by Fig. \ref{Fig:FactorLSCS} and Fig. \ref{Fig:FactorCSCS}:
\begin{itemize}
    \item \textbf{The effective factors in one row are limited.} For example, in the first row, the effective factors are only $C_1I+S_1Z$, $C_2I+S_2Z$, $C_{N+2}I+S_{N+2}Z$, $C_{N+2}I+S_{N+2}Z$, $C_{N+3}I+S_{N+3}Z$, and $C_{1'}U+S_{1'}D$ (the matrices in other rows are not revealed). The rest of the bracketed factors in this row have the form $CI+SI$, which are composed only of identities. These bracketed factors do not change $I$s or $Z$s in other bracketed factors. The complex numbers $C$ and $S$ can be moved to other matrices in the column of the tensor products. This ensures that one can compute the multiplications in one row with bounded steps.
    \item \textbf{Most of the matrices in a tensor are identities. As a consequence, the effective connections of the effective factors to those in the next row are also limited, and smaller than the number of effective factors in one row. As such, in the recursive relations, the result of one step is only a function of the limited results obtained in the previous step.} For example, in the first and second rows, the effective connections are included in the factors $C_2(I\otimes I)+S_2(Z\otimes Z)$ and $C_{N+3}(I\otimes I)+S_{N+3}(Z\otimes Z)$ (the matrices in the other rows are also not revealed). The other factors involving two rows are $(C_1I+S_1Z)\otimes I$, $(C_{N+2}I+S_{N+2}Z)\otimes I$, and $(C_{1'}U+S_{1'}D)\otimes(C_{2'}U+S_{2'}D)$, which are pure tensor products. This ensures that the multiplication result of one row only affects small number factors in the next row, so that the recursive relation in one step is only a function of limited outcomes obtained from the previous calculations (such as Eq. (\ref{eq:LSCS-C4}) and Eq. (\ref{eq:CSCS-C1}), which contains only two terms of the previous computation results). 
    \item \textbf{The total number of rows is in the first-order relation with the total number of qubits.} This is relatively easy to notice. The total number of rows equals the number of target qubits (blue dots), which is $MN$.
\end{itemize}
Hence, one might consider that the local projections on the 2-D extension of the cross-shape cluster states in Fig. \ref{Fig:CSCS}, or equivalently the 2DSCLSs in Fig. \ref{Fig:2DCS}, can be computed in $MN$ steps with bounded memory. Again, $MN$ is in the first order of the total number of qubits. Also, when using the recursive relations, it is not necessary to restore all previous computation results other than those obtained in one step before the current, leading to a bounded cost of memory. 

\begin{figure}
\centering
\includegraphics[width=3.4in]{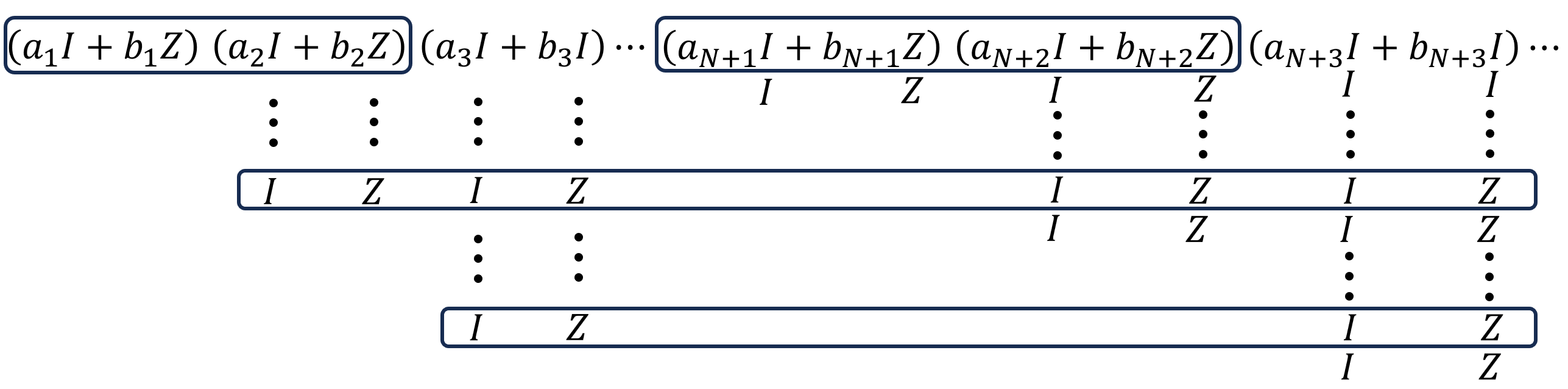}
\caption{A form of the polynomial in Fig. \ref{Fig:2DLSfac} after re-ordering the brackets. The parameters $C$ and $S$ have been changed to $a$ and $b$. This is a lazy treatment, because the authors would not like to deduce the subscripts of $C$ and $S$ after the re-ordering. The black boxes also mark the matrices that needs to be specially noticed in the multiplication.}\label{Fig:Refac-1}
\end{figure}

\begin{figure}
\centering
\includegraphics[width=3.4in]{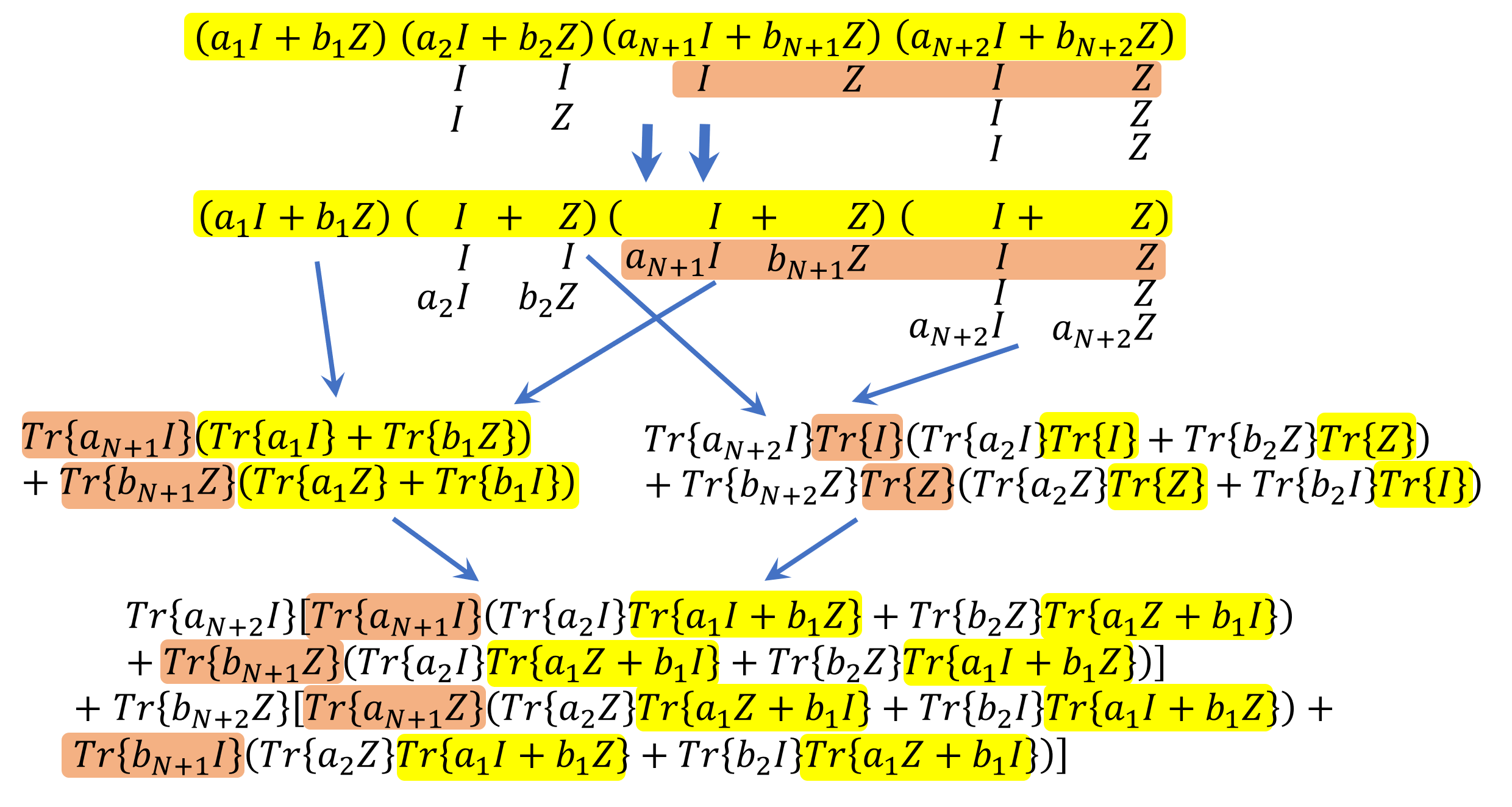}
\caption{The process of distributing the terms across four brackets involving $a_1$, $b_1$, $a_2$, $b_2$, $a_{N+1}$, $b_{N+1}$, $a_{N+2}$, and $b_{N+2}$. The upper double arrows denote the procedure of shifting the complex numbers. The lower arrows denote the procedure of obtaining the traces. By comparing the changes of the traces, one might find the patterns. The terms affected by the same bracketed factors are colored the same (yellow or red).}\label{Fig:Refac-2}
\end{figure}

\begin{figure}
\centering
\includegraphics[width=3.4in]{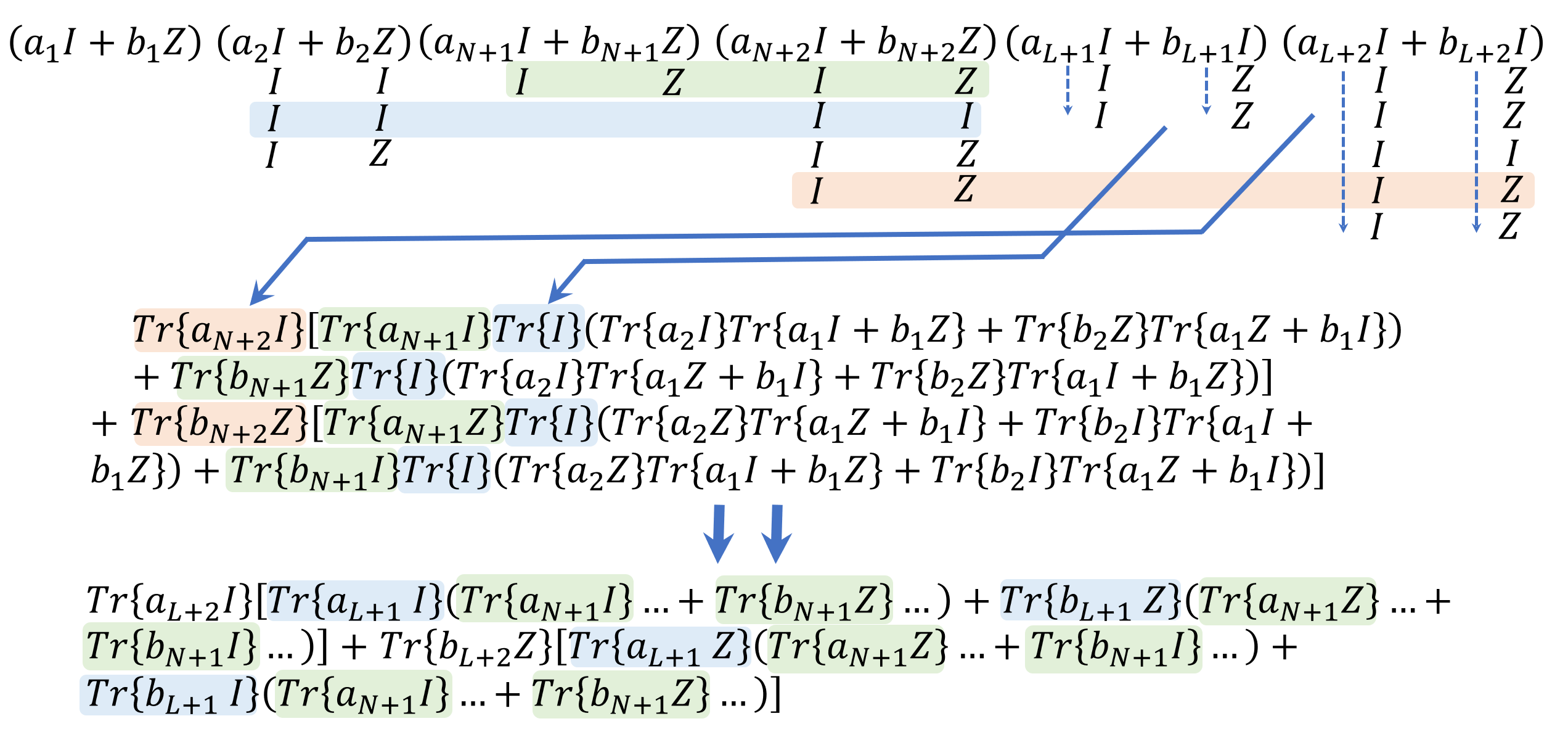}
\caption{One way of distributing the terms across the brackets, when considering two more bracketed factors than the case in Fig. \ref{Fig:Refac-2}. This leads to enlarging the size of the tensor product, from $\{I\otimes I,~Z\otimes I,~Z\otimes Z\}$ to $\{I\otimes I\otimes I,~Z\otimes I\otimes I, Z\otimes Z\otimes I,~I\otimes Z\otimes Z\}$ (the ``tensor square terms'' in Eq. (\ref{eq:Refac-3}) like $(I\otimes I)^{\otimes2}$ or $(Z\otimes Z)^{\otimes2}$ are similar, but not shown). The two arrows with curved lines in the upper place mark the terms that are affected by the two added bracketed factors. The double arrows in the lower place mark the results of the affection, in which several irrelevant terms are omitted. The terms affected by the same bracketed factors are also colored the same (pink, blue, or green).}\label{Fig:Refac-3}
\end{figure}

\begin{figure}
\centering
\includegraphics[width=3.4in]{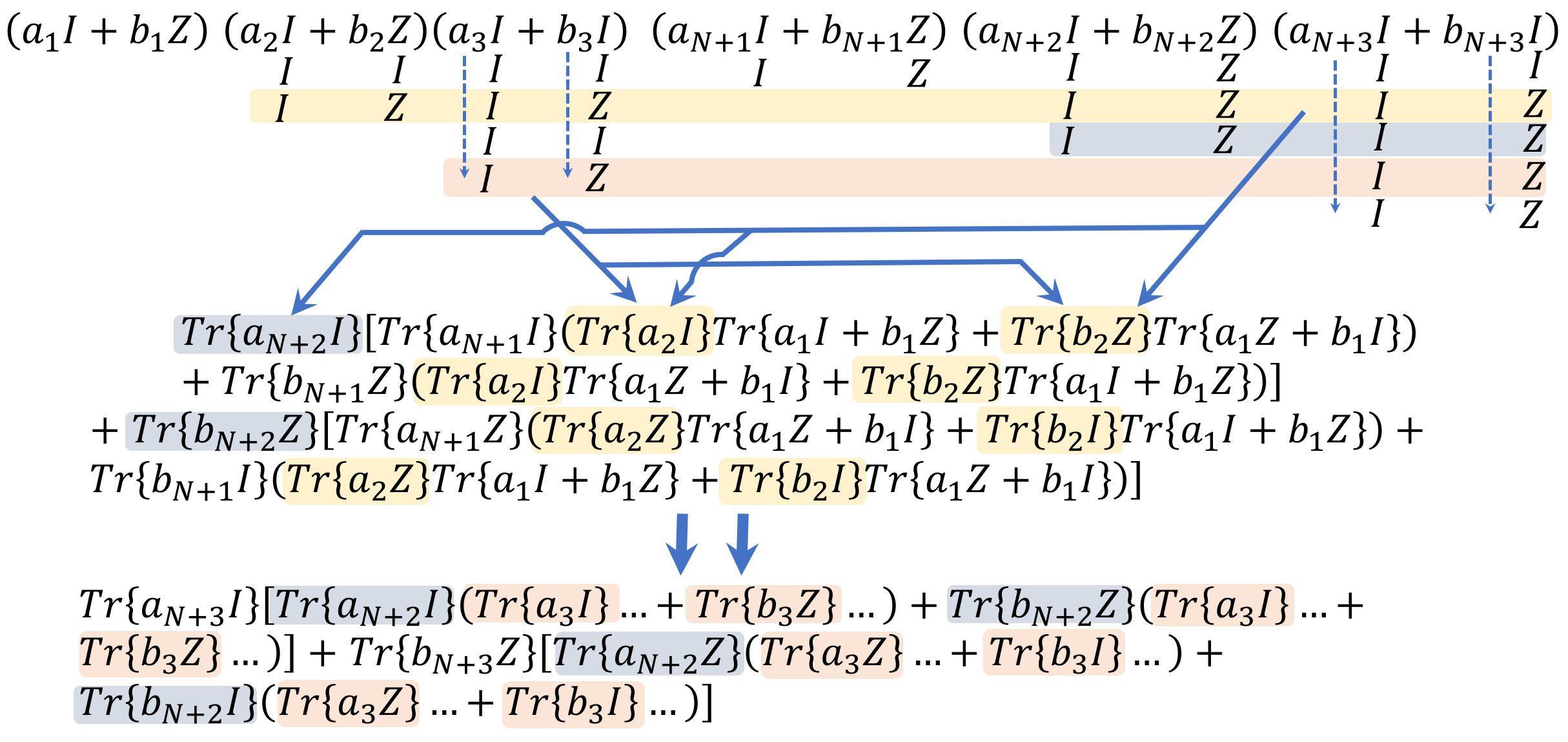}
\caption{Another way of distributing the terms across the brackets, when considering two more factors than the case in Fig. \ref{Fig:Refac-2}. This also leads to enlarging the size of the tensor product, but in a different way. The tensor form changes from $\{\{I\otimes I,~Z\otimes Z\}\otimes I^{\otimes 2}\}\cup\{~(I\otimes I)^{\otimes2},~(Z\otimes Z)^{\otimes2}\}$ to $\{\{I\otimes I,~Z\otimes Z\}\otimes I^{\otimes 4}\}\cup\{\{~(I\otimes I)^{\otimes2},~(Z\otimes Z)^{\otimes2}\}\otimes I^{\otimes 2}\}\cup\{I^{\otimes 2}\otimes\{~(I\otimes I)^{\otimes2},~(Z\otimes Z)^{\otimes2}\}\}$ (the author uses $\{\{I\otimes I,~Z\otimes Z\}\otimes I^{\otimes 2}\}$ as a shorten form for $\{I\otimes I\otimes I^{\otimes 2},~Z\otimes Z\otimes I^{\otimes 2}\}$, and other similar forms follow the same logic). The terms affected by the two added bracketed factors are also marked out by the arrows with curved lines in the upper place. The results of the affection are also indicated by the double arrows in the lower place, with several irrelevant terms being omitted. The terms affected by the same bracketed factors are also colored the same as well (pink, orange, or gray).}\label{Fig:Refac-4}
\end{figure}

Of course, the above conclusion is NOT a strict one. A fluctuation lies in the second character listed above. It can be seen that, other than the effective connections to the bracketed factors in the next row, there are also effective connections to the factors in the row of other blocks. This fact causes some troubles for obtaining the recursive relations. However, we believe that it does not destroy the above strategy. This can be seen from the step-by-step procedure shown in Figs. \ref{Fig:Refac-1}-\ref{Fig:Refac-4}. 

In Fig. \ref{Fig:Refac-1}, a polynomial is given. It is obtained by reordering the bracketed factors of the polynomial in Fig. \ref{Fig:2DLSfac}. As explained in the caption, the author does not provide an explicit relation between the parameters $C$ and $S$ in the previous expression and the parameters $a$ and $b$ in Fig. \ref{Fig:Refac-1}. It is just a reordering of the factors, so one can formally have the expression. As shown above, the recursive relations can be obtained by distributing the terms across the brackets. Observing the expression in Fig. \ref{Fig:Refac-1} , it can be found that the distribution of the terms can be carried out in two ``directions''. In Figs. \ref{Fig:Refac-2}, \ref{Fig:Refac-3}, and \ref{Fig:Refac-4}, the meaning of the directions are graphically illustrated. 

In the first step, one can set a starting point by taking four bracketed factors, as shown in Fig. \ref{Fig:Refac-2}. Here, the factors involving $a_1$, $b_1$, $a_2$, $b_2$, $a_{N+1}$, $b_{N+1}$, $a_{N+2}$ are chosen, because their effective matrices are in the front rows. As in the previous notations, the column of $I$ and $Z$ denotes the tensor product of them. Therefore, the complex numbers $a$ and $b$ can be shifted to the front of other matrices other than those in the first row, marked by the blue arrows in the upper place. Then, to show the form, one can first multiply the bracket involving $a_1$ and $b_1$ with the bracket involving $a_{N+1}$ and $b_{N+1}$, secondly multiply the bracket involving $a_2$ and $b_2$ with the bracket involving $a_{N+2}$ and $b_{N+2}$, and thirdly take the trace of them. Afterward, by distributing all four brackets and taking the trace, one might notice that the result of the total expression (shown at the bottom of Fig. \ref{Fig:Refac-2}) can be generated from the expressions of the traces given by the products of the two brackets (shown in the middle of Fig. \ref{Fig:Refac-2}). Because the terms affected by the same factors are colored the same, as shown in Fig. \ref{Fig:Refac-2}, the generation can be done by replacing the terms on the right by the terms of the same color on the left. When dealing with multiple choices of replacements, one just simply copies the terms to be replaced and replaces them with all different choices. 

In the second step, consider taking more bracketed factors into account and observe the changes of the trace form. This gives the ``two directions'' mentioned earlier. In one of the directions, the number of matrices in a block increases. Inducing such changes, one can observe the transformation of the projection polynomial along the rows of the graph in Fig. \ref{Fig:2DLSExten}. Particularly, it is given by 
\begin{equation}\label{eq:Refac-3}
    \begin{split}
    &(a_1I\otimes I\otimes I^{\otimes2}+b_1Z\otimes I\otimes I^{\otimes2})(a_2I\otimes I\otimes I^{\otimes2}\\
    +&b_2Z\otimes Z\otimes I^{\otimes2})[a_{N+1}(I\otimes I)^{\otimes2}+b_{N+1}(I\otimes Z)^{\otimes2}]\\
    \cdot&[a_{N+2}(I\otimes I)^{\otimes2}+b_{N+2}(Z\otimes Z)^{\otimes2}]\\
    \rightarrow&(a_1I\otimes I\otimes I\otimes I^{\otimes 3}+b_1Z\otimes I\otimes I\otimes I^{\otimes 3})(a_2I\otimes I\otimes I\\
    &\otimes I^{\otimes 3}+b_2Z\otimes Z\otimes I\otimes I^{\otimes 3})(a_{L+1}I\otimes I\otimes I\otimes I^{\otimes 3}\\
    +&b_{L+1}I\otimes Z\otimes Z\otimes I^{\otimes 3})[a_{N+1}(I\otimes I\otimes I)^{\otimes2}+b_{N+1}(Z\\
    &\otimes I\otimes I)^{\otimes2}][a_{N+2}(I\otimes I\otimes I)^{\otimes2}+b_{N+2}(Z\otimes Z\\
    &\otimes I)^{\otimes2}][a_{L+2}(I\otimes I\otimes I)^{\otimes2}+b_{L+2}(I\otimes Z\otimes Z)^{\otimes2}].
    \end{split}
\end{equation}
This extension is graphically shown in Fig. \ref{Fig:Refac-3}, and several replicas of $I$ are omitted---it is perhaps easier to read than the expression (\ref{eq:Refac-3}). By analyzing the changes of the trace after the extension, one might notice that only partial terms are transformed. In Fig. \ref{Fig:Refac-3}, the terms affected by the same bracketed factors are also colored the same, and the main affections of the two added bracketed factors are specially marked by the arrows with curved tails in the upper place. The resultant form is given in the lower place, marked by the double arrows. It is like the pattern shown in Fig. \ref{Fig:Refac-2}, but more complicated. 

In the other direction, the number of blocks increases, but the number of matrices remains the same. Inducing such changes, one can observe the transformation of the projection polynomial along the columns of the graph in Fig. \ref{Fig:2DLSExten}. It is particularly given by, 
\begin{equation}\label{eq:Refac-4}
    \begin{split}
    &(a_1I\otimes I\otimes I^{\otimes2}+b_1Z\otimes I\otimes I^{\otimes2})(a_2I\otimes I\otimes I^{\otimes2}\\
    +&b_2Z\otimes Z\otimes I^{\otimes2})[a_{N+1}(I\otimes I)^{\otimes2}+b_{N+1}(I\otimes Z)^{\otimes2}]\\
    \cdot&[a_{N+2}(I\otimes I)^{\otimes2}+b_{N+2}(Z\otimes Z)^{\otimes2}]\\
    \rightarrow&(a_1I\otimes I\otimes I^{\otimes 4}+b_1Z\otimes I\otimes I^{\otimes 4})(a_2I\otimes I\otimes I^{\otimes 4}\\
    +&b_2Z\otimes Z\otimes I^{\otimes 4})[a_{N+1}(I\otimes I)^{\otimes2}\otimes I^{\otimes 2}+b_{N+1}(Z\\
    \otimes& I)^{\otimes2}\otimes I^{\otimes 2}][a_{N+2}(I\otimes I)^{\otimes2}\otimes I^{\otimes 2}+b_{N+2}(Z\\
    \otimes& Z)^{\otimes2}\otimes I^{\otimes 2}][a_3I^{\otimes 2}\otimes(I\otimes I)^{\otimes2}+b_3I^{\otimes 2}\otimes(I\\
    \otimes& Z)^{\otimes2}][a_{N+3}I^{\otimes 2}\otimes(I\otimes I)^{\otimes2}+b_{N+3}I^{\otimes 2}\otimes(Z\\
    \otimes& Z)^{\otimes2}].
    \end{split}
\end{equation}
Similarly to the previous direction, this extension is shown graphically in Fig. \ref{Fig:Refac-4}, and the replicas of $I$ are also omitted. By observing the changes of the trace after the extension, again, one might also notice the limit in the number of terms that are transformed. In Fig. \ref{Fig:Refac-4}, like the above, the terms affected by the same bracketed factors are also colored the same, and the main affections of the two added bracketed factors are specially marked by the arrows with curved tails in the upper place. The resultant form is given in the lower place, marked by the double arrows. This also demonstrates the pattern shown in Fig. \ref{Fig:Refac-2}, but is more complicated as well.

In all, for the polynomial shown in Fig. \ref{Fig:2DLSfac}, one may establish recursive relations by using the transformations deduced from the extensions in Figs. \ref{Fig:Refac-3} and \ref{Fig:Refac-4}. For ease of action, one might consider expanding the brackets along the anti-diagonal direction of Fig. \ref{Fig:2DLSExten}, that is, $1\rightarrow (N+2)\rightarrow 2\rightarrow 3\rightarrow (N+3)\rightarrow(2N+3)\rightarrow\cdots$. In fact, such a pattern can also be drawn by directly observing the symmetry of the polynomial in Fig. \ref{Fig:2DLSfac}, or even the original local projections on the 2DSLCS. The graphic illustration given by Figs. \ref{Fig:2DLSfac} to \ref{Fig:Refac-4} are only used to find the pattern. 

Next, a different view of the polynomial in Fig. \ref{Fig:2DLSfac} is provided. Before we discuss the details of the view, let us first consider a special set. It can be noticed that the exponential set $\{I,Z\}^{\otimes N}=\{I\otimes\cdots\otimes I,~Z\otimes\cdots\otimes I,\ldots,~Z\otimes\cdots\otimes Z\}$ can be considered as a basis set of a linear space on the complex number field, defined by 
\begin{equation}
    \begin{split}
    &\mathfrak{span}(\{I,Z\}^{\otimes N})=\{c_1I\otimes\cdots\otimes I+c_2Z\otimes\cdots\otimes I+\\
    &~~~~~~~~~~~~~~~\cdots+c_{2^N}Z\otimes\cdots\otimes Z|c_1,\ldots,c_{2^N}\in\mathbb{C}\}.   
    \end{split}
\end{equation}
It is easy to check the following two properties. First, the linearity. The complex linear combinations of the elements in $\mathfrak{span}(\{I,Z\}^{\otimes N})$ are still in the set. Second, the closure property in matrix multiplication. The matrix multiplication of any two elements in $\mathfrak{span}(\{I,Z\}^{\otimes N})$ is still in the set. In addition, the $2^N$-dimensional zero matrix and the identity matrix are also in $\mathfrak{span}(\{I,Z\}^{\otimes N})$, which can be viewed as the zero and identity elements of the set. Hence, such a set is a ring.

Following the above definition, the factors in Fig. \ref{Fig:2DLSfac}, such as $(C_1I+S_1Z)\otimes I\cdots$ or $(C_1I\otimes I+S_1Z\otimes Z)\otimes I\cdots$, are in $\mathfrak{span}(\{I,Z\}^{\otimes MN})$. Thus, one may consider expanding the brackets in blocks. Starting from the first column of blocks, which has $N+1$ brackets of factors (it equals to the number of the qubits in the first row of the graph in Fig. \ref{Fig:2DLSExten}), the expansion is given by 
\begin{equation}\label{eq:blockexpan-1}
    \begin{split}
    &(\sigma_1 I\otimes\cdots\otimes I+\sigma_2 Z\otimes I\cdots\otimes I+\cdots+\sigma_{2^{N+1}}Z\\
    &\otimes\cdots\otimes Z)\otimes I\otimes I\cdots.   
    \end{split}
\end{equation}
Obviously, $\sigma_1$ to $\sigma_{2^{N+1}}$ can be given by the parameters $C$ and $S$. Then, by using the relation $U=(I+Z)/2$ and $U=(I-Z)/2$, the factors $(C_{1'}U+S_{1'}D)\otimes(C_{1'}U+S_{1'}D)\otimes\cdots\otimes(C_{N'}U+S_{N'}D)$ can be expressed by the element in $\mathfrak{span}(\{I,Z\}^{\otimes N})$. Afterward, if one multiplies expression (\ref{eq:blockexpan-1}) and above tensor product involving $U$ and $D$, the obtained polynomial shares the same basic form as expression (\ref{eq:blockexpan-1}), and only the parameters are different. One may consider the transformation $\sigma\rightarrow\tilde{\sigma}$. For the second column of the blocks, which also has $N+1$ brackets of factors, the expansion is given by
\begin{equation}\label{eq:blockexpan-2}
    \begin{split}
    &[\delta_1 (I\otimes\cdots\otimes I)^{\otimes2}+\delta_2(Z\otimes I\cdots\otimes I)^{\otimes2}+\cdots\\
    &+\delta_{2^{N+1}}(Z\otimes\cdots\otimes Z)^{\otimes2}]\otimes I\otimes I\cdots.   
    \end{split}
\end{equation}
Notice that the multiplication of expression (\ref{eq:blockexpan-1}) and (\ref{eq:blockexpan-2}) will lead to a polynomial with two characters. For the first, expression (\ref{eq:blockexpan-1}) only affects ``half'' of the tensor products in the square bracket of expression (\ref{eq:blockexpan-1}). For the second, the non-zero component of the trace has a pattern. An easy example would help to make the pattern clearer. Consider the case when $N=1$ (this actually goes back to the case in Fig. \ref{Fig:CSCS}, but it still helps to see the case here), one has 
\begin{equation}\label{eq:blockexpan-3}
\begin{split}
    \mathcal{M}_1&=(\tilde{\sigma}_1 I\otimes I+\tilde{\sigma}_2 Z\otimes I+\tilde{\sigma}_3 I\otimes Z\\
    &~~+\tilde{\sigma}_{4}Z\otimes Z)\otimes I\otimes I,
\end{split}
\end{equation}
and
\begin{equation}\label{eq:blockexpan-4}
    \begin{split}
    \mathcal{M}_2&=\delta_1 (I\otimes I)^{\otimes2}+\delta_2(Z\otimes I)^{\otimes2}+\delta _3(I\otimes Z)^{\otimes 3}\\
    &~~+\delta_4(Z\otimes Z)^{\otimes2}.    
    \end{split}
\end{equation}
The replicas of $I$ in the tensor products are omitted, because they do not contribute much to the trace. In addition, due to the fact that $Tr\{Z\}=0$, once the tensor product has one $Z$, the trace of it will be $0$. Therefore, taking the trace of the product of expression (\ref{eq:blockexpan-3}) and (\ref{eq:blockexpan-4}) results in
\begin{equation}\label{eq:blockexpan-5}
\begin{split}
    &Tr\{\mathcal{M}_1\mathcal{M}_2\}\\
    =&Tr\{\delta_1 (I\otimes I)^{\otimes2}\mathcal{M}_1\}+Tr\{\delta_2(Z\otimes I)^{\otimes2}\mathcal{M}_1\}\\
    +&Tr\{\delta _3(I\otimes Z)^{\otimes2}\mathcal{M}_1\}+Tr\{\delta_4(Z\otimes Z)^{\otimes2}\mathcal{M}_1\}\\
    =&4\cdot Tr\{\delta_1\tilde{\sigma}_1I\otimes I\}+4\cdot Tr\{\delta_2\tilde{\sigma}_2Z\otimes I\}\\
    +&4\cdot Tr\{\delta_3\tilde{\sigma}_3I\otimes Z\}+4\cdot Tr\{\delta_4\tilde{\sigma}_4Z\otimes Z\}.
\end{split}
\end{equation}
The pattern mentioned above is mainly shown by the last equation in Eqs. (\ref{eq:blockexpan-5}). Because no $Z$ would be left if and only if two tensors of the same form are multiplied, such as $(I\otimes I)(I\otimes I)$ or $(I\otimes Z)(I\otimes Z)$, many terms in the trace are zeros. For example, in the trace $Tr\{(I\otimes Z)^{\otimes 2}\mathcal{M}_1\}$, only the parameter $\tilde{\sigma}_3$ of $I\otimes Z\otimes I\otimes I$ in $\mathcal{M}_1$ is important and left, and the rest of the parameters are involved in the tensors with zero trace. Therefore, by simply multiplying the parameters $\tilde{\sigma}_i$ and $\delta_i$ ($i=1,\ldots,4$), the blocks in the original polynomial can be reduced, from like $(Z\otimes I)^{\otimes2}$ or $Z\otimes I^{\otimes3}$ to $Z\otimes I$. Furthermore, such a trick can be applied again and again to the computation of the trace in Fig. \ref{Fig:2DLSfac}, so that the final trace is proportional to 
\begin{equation}\label{eq:totalexpan-1}
    \begin{split}
    &Tr\{\Pi_1 I\otimes\cdots\otimes I+\Pi_2 Z\otimes I\cdots\otimes I+\cdots+\Pi_{2^{N+1}}Z\\
    &\otimes\cdots\otimes Z\}, 
    \end{split}
\end{equation}
where $\Pi_i$ ($i=1,\ldots,2^{N+1}$) are the products of parameters in the expansions of the column of blocks, such as $\sigma$ and $\delta$ in expression (\ref{eq:blockexpan-1}) and (\ref{eq:blockexpan-2}). In fact, expression (\ref{eq:totalexpan-1}) equals to $2^{N+1}\Pi_1$. From Fig. \ref{Fig:2DLSfac}, it can be noticed that there would be $M+1$ columns of the blocks, which generates $M+1$ expressions after the expansion. Those expressions are of the similar forms like expressions (\ref{eq:blockexpan-1}) and (\ref{eq:blockexpan-2}). Although there are $2^{N+1}$ terms in each of the $M+1$ expressions, the computation of the parameters such as $\delta_1$ or $\tilde{\sigma}_1$ can be done in polynomial steps. This is already verified by the results in Section \ref{subsec:CSCS}. In all, the above pattern leads to a polynomial strategy for computing the projection on a 2DSLCS. 

Another view is also provided on the polynomial given in Fig. \ref{Fig:2DLSfac}. Observing the form, one might think that the tensor product $(C_{1'}U+S_{1'}D)\otimes (C_{2'}U+S_{2'}D)\cdots$ looks like a tensor product of mixed state density matrices. Meanwhile, the factor $(C_aI+S_aZ)$ or $(C_aI\otimes I+S_aZ\otimes Z)$ (with integer $a$) looks like a Pauli-$Z$ rotation considering the exponential form $e^{i\alpha Z}$ or $e^{i\alpha Z\otimes Z}$. Hence, the polynomial in Fig. \ref{Fig:2DLSfac} seemingly corresponds to a special quantum circuit, which is composed of Pauli-$Z$ rotations and $ZZ$ Ising gates and employs mixed state as the input. However, in standard quantum mechanics, the output state after evolution is given by $\mathcal{U}_C^{\dagger}\rho \mathcal{U}_C$, in which $\mathcal{U}_C$ is the circuit operator and $\rho$ is the density matrix of the mixed state. Compared to that, the polynomial in Fig. \ref{Fig:2DLSfac} is more like $\mathcal{U}_C\rho$, which is only traced to a complex number. 

\section{Several Particular Examples with finite qubits and numerical estimations}\label{sec:egs}
In this section, several particular examples are provided, further validating the above trick. 

\begin{figure*}
\centering
\includegraphics[width=6.7in]{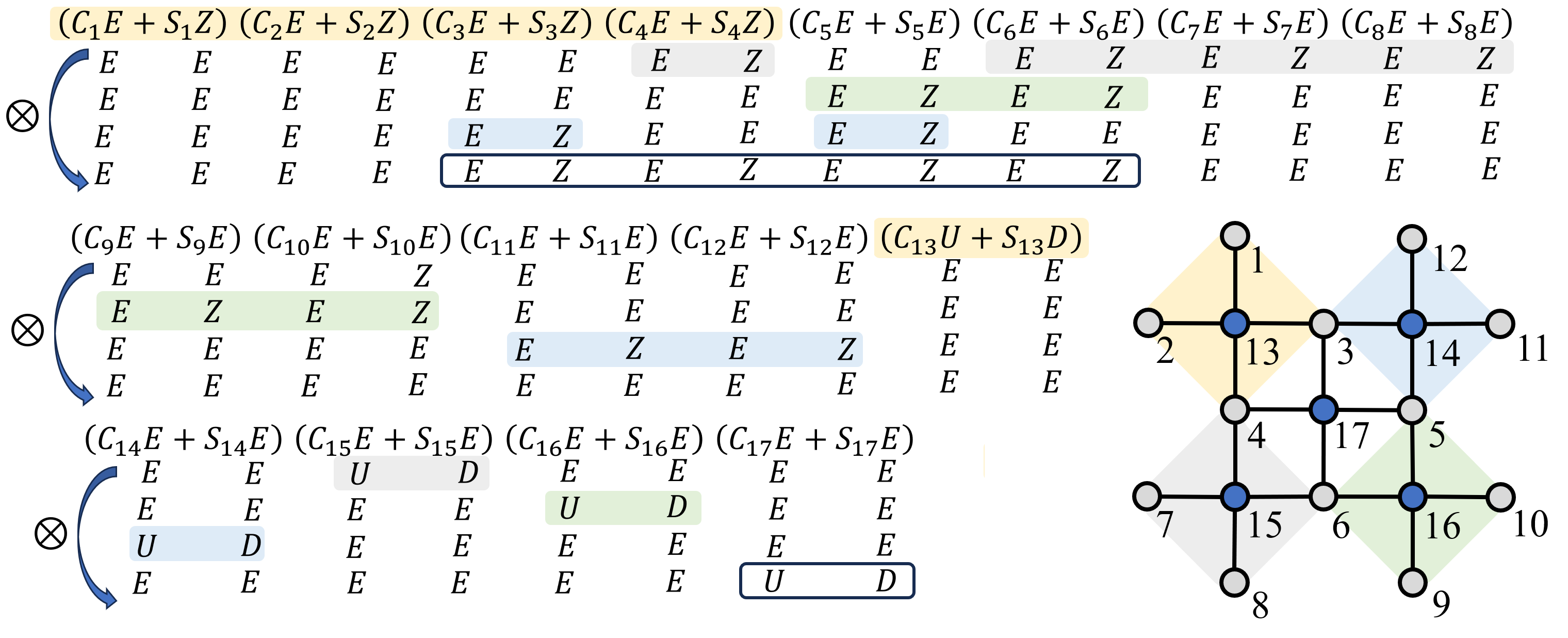}
\caption{The factorization of projection on a finite 2DSLCS in the right corner of the bottom. The matrices $E$ (equals to $I$), $Z$, $U$, and $D$ in each row are multiplied under tensor product rules. The brackets of the factors with colored background in the expression describe the region of the graph in the right-down corner that is colored the same. The ones in the black box describe the middle cross-shape state in the graph of the total state.}\label{Fig:Fact.Eg.}
\end{figure*}

In the first example, a cluster state composed of five cross-shape cluster states is considered, as shown in the right corner of the bottom of Fig. \ref{Fig:Fact.Eg.}. It is also a 2DSLCS, but in a different composition method than the one in Fig. \ref{Fig:2DLSExten}. Of course, the two composition methods are equivalent to each other. The factorization of the projection on the state can also be given by following the strategy in the previous discussion, and it is shown in the rest of the figure. There are 17 brackets in the polynomial in total, corresponding to the 17 qubits in the cluster state. The figure is drawn when drafting the early version of the note, so the 2-by-2 identity matrix is denoted by $E$ rather than $I$. The reader only needs to know that the matrix $E$ in Fig. \ref{Fig:Fact.Eg.} equals $I$ in the previous context. The brackets of factors used to describe the qubits in the same cross-shape cluster state are colored the same. For example, the cross-shape cluster state composed of the 1st, the 2nd, the 3rd, the 4th, and the 13th qubits corresponds to the product of the brackets $(C_1E+S_1Z)(C_2E+S_2Z)(C_3E+S_4Z)(C_4E+S_4Z)(C_{13}U+S_{13}D)\otimes E^{\otimes 4}$. They are both colored yellow. The rest of the qubits and the bracketed factors are similar. The bracketed factors for describing the cross-shape cluster state in the middle (composed of the 3rd, the 4th, the 5th, the 6th, and the 17th qubits) are marked by black boxes. In fact, it is easy to check that the trace of the polynomial is proportional to the projection of the cluster state in the figure. The constant factor, which is actually $2^{-17/2}$, is also easy to compute by normalizing the final result of the trace. Notice that a direct strategy for computing the projection of the cluster state in Fig. \ref{Fig:Fact.Eg.} usually requires to generate $2^{17}$-order matrices. However, the trace of the factorized form only requires one to generate at most $2^{5}$-order matrices, let alone the application of the recursive relations for this particular case. We think of it as an interesting simplification strategy. 

\begin{figure}
\centering
\includegraphics[width=3.4in]{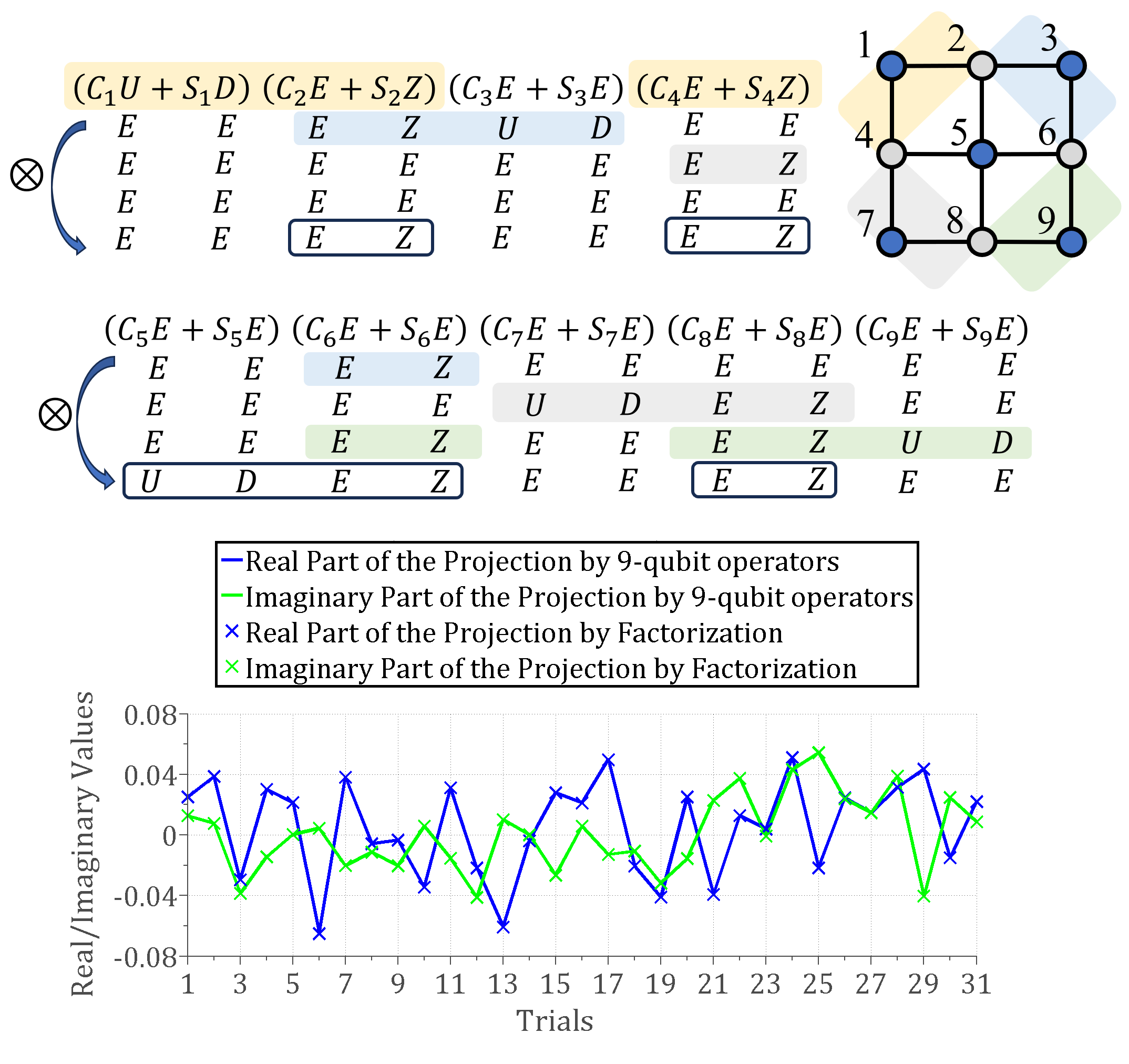}
\caption{The factorization form of the projection on a 3-by-3 2DSLCS using the above strategy is given in the upper panel. Similar to the case in Fig. \ref{Fig:Fact.Eg.}, the matrices $E$, $Z$, $U$, and $D$ in each row are multiplied under the rule of tensor product. Also, the terms with colored background in the expression describe the region of the graph in the right-down corner that is colored the same. The numerical results is given in the lower panel. The curves represent the results obtained by directly computing the inner products. The dots with cross marks represent the results obtained by computing the factorized polynomial in the above. The blue and the green ones represent the real and imaginary parts of the results.}\label{Fig:Fact.Eg.2}
\end{figure}

In the second example, a numerical validation of a smaller 2DSLCS is given, and the results are illustrated in Fig. \ref{Fig:Fact.Eg.2}. The cluster state in the example is a 3-by-3 2DSLCS, as shown in the right corner of the top. The factorization of the projection on the state is given next to the graph of the state. Similarly to the first example in this section, Fig. \ref{Fig:Fact.Eg.2} is also drawn for the early version, so notation $E$ is used instead of $I$. The numerical evaluations of the projection by different strategies are presented below. In the plot, the curves are obtained by directly computing the inner products of the projection vectors and the cluster state vectors, which are both $2^9$-dimensional complex vectors. The dots with cross marks are obtained by computing the trace of the matrix polynomial, which involves only the calculation of $2^5$-dimensional complex vectors. Because this case is quite simple, recursive relations are not derived.  

For plotting the curves in the lower panel, 31 random trails ($\theta$ and $\varphi$ in $C$ and $S$ are randomly chosen from $0$ to $2\pi$) are performed. The blue and green ones represent the real and imaginary parts of the results. From Fig. \ref{Fig:Fact.Eg.2}, it is easy to see that the two methods generate the same results (this needs to consider the factor $2^{-9/2}$ which is not shown in the polynomial). The running time of the methods are significantly different. Without optimization, the average running time of the computation by inner product is $0.0995$ seconds, which is about $0.1$ seconds. The average running time of the computation by tracing the factorized polynomial is $10^{-3}+2.4658\times10^{-4}$ seconds, where $10^{-3}$ seconds is the time for computing the tensor products in the expression (such as $U\otimes E\otimes E\otimes E\otimes E$, etc.) and $2.4658\times10^{-4}$ seconds is the time for computing the linear combination of the tensor products and their multiplications. Considering that those tensor products are fixed once the connections of the 2DSLCS is given, they can be computed in advance. This will increase the consumption in memory and guarantee the ruining time of $2.4658\times10^{-4}$ seconds, which is about three orders lower than the original inner product method. The computational resource is given as follows: 1) the CPU is Intel (R) Core (TM) i7-7700, with main frequency 3.60 GHz; 2) the size of the RAM is 8 GB; and 3) the operating system is windows 10 (based on x64 processor). The calculation software is Matlab (ver. 2012b).

\begin{figure}
\centering
\includegraphics[width=3.4in]{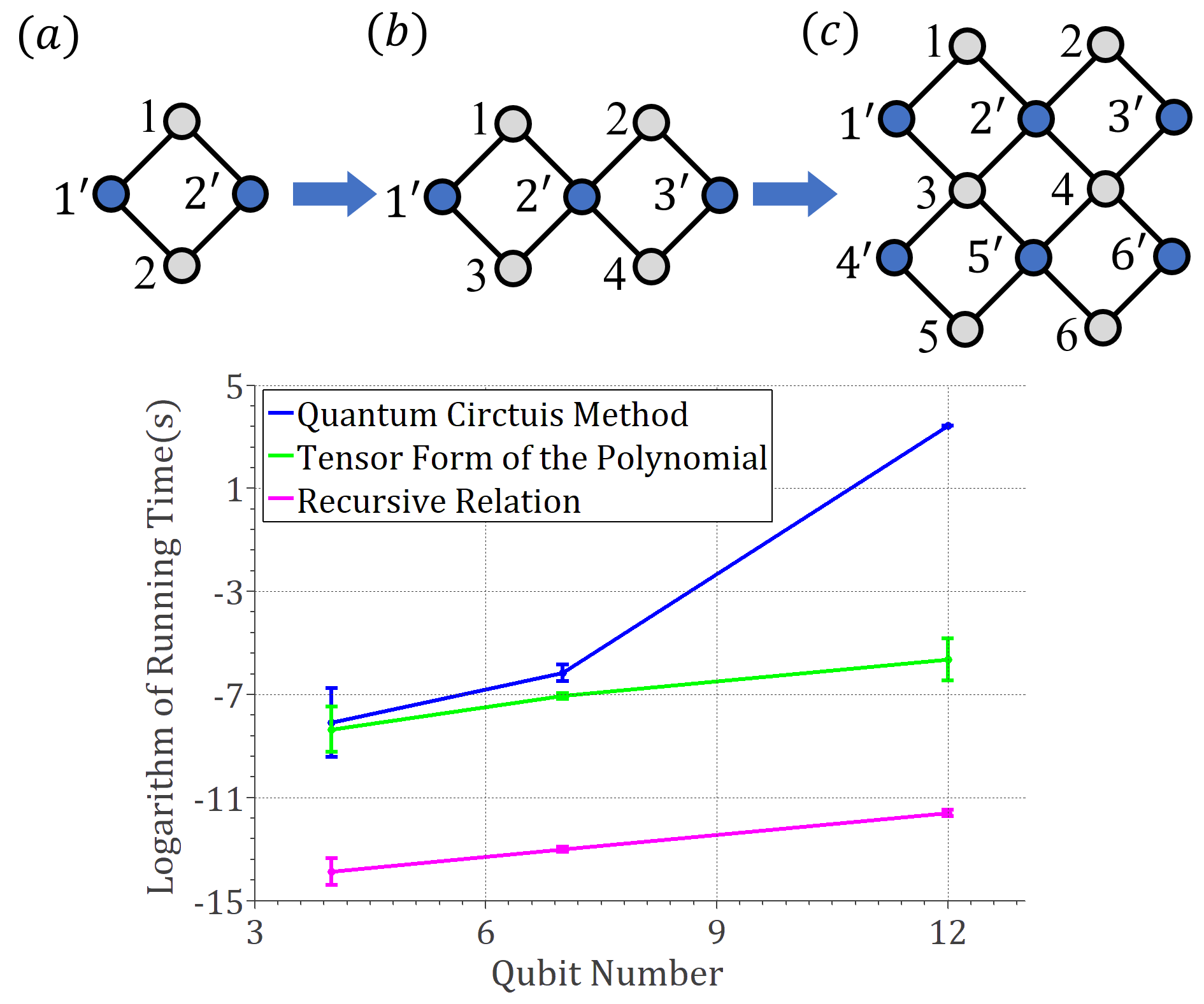}
\caption{The running time of the three algorithms for computing the projections on the cluster states shown in (a), (b), and (c). The states in (a), (b), and (c) are composted of 4, 7, and 12 qubits respectively. The blue curve is obtained by generating the complex state vectors, and project them onto a random direction. This is noted by ``quantum circuits method'', and the circuits for generating the complex state vectors are shown in Fig. \ref{Fig:Fact.Eg.3-1}. The green and the pink curves are obtained by the factorized polynomial, which are shown in Fig. \ref{Fig:Fact.Eg.3-2}. Particularly, the green curve is obtained by computing the trace of the tensor form of the polynomial, and the pink curve is obtained by using the recursive relations derived from the polynomial. The value of each point in the curves is an average over 25 times of trials, and the standard deviations are also marked on the points. The projection directions of the state in one trial for the three algorithms are the same and randomly chosen. The running time for computing one state are obtained when the outputs of the three algorithms are completely the same.}\label{Fig:Fact.Eg.3}
\end{figure}

In the third example, a numerical estimation of the scaling is given. Consider three cluster states, which are composed of 4, 7, 12 qubits, respectively, and are shown at the top of Fig. \ref{Fig:Fact.Eg.3}. Despite that those in (a) and (b) are quite simple, they can both be viewed as 2DSLCSs roughly. In the lower panel, the average running time of projection by different methods are given. For ease of visualization, the natural logarithm of the running time is used as the vertical coordinates of the data in the plot. All data is obtained under the condition that the outputs of all three methods for computing the same projections, both real parts and imaginary parts, are completely the same as each other. The computational environment is the same as that in the second example in this section.

\begin{figure}
\centering
\includegraphics[width=3.4in]{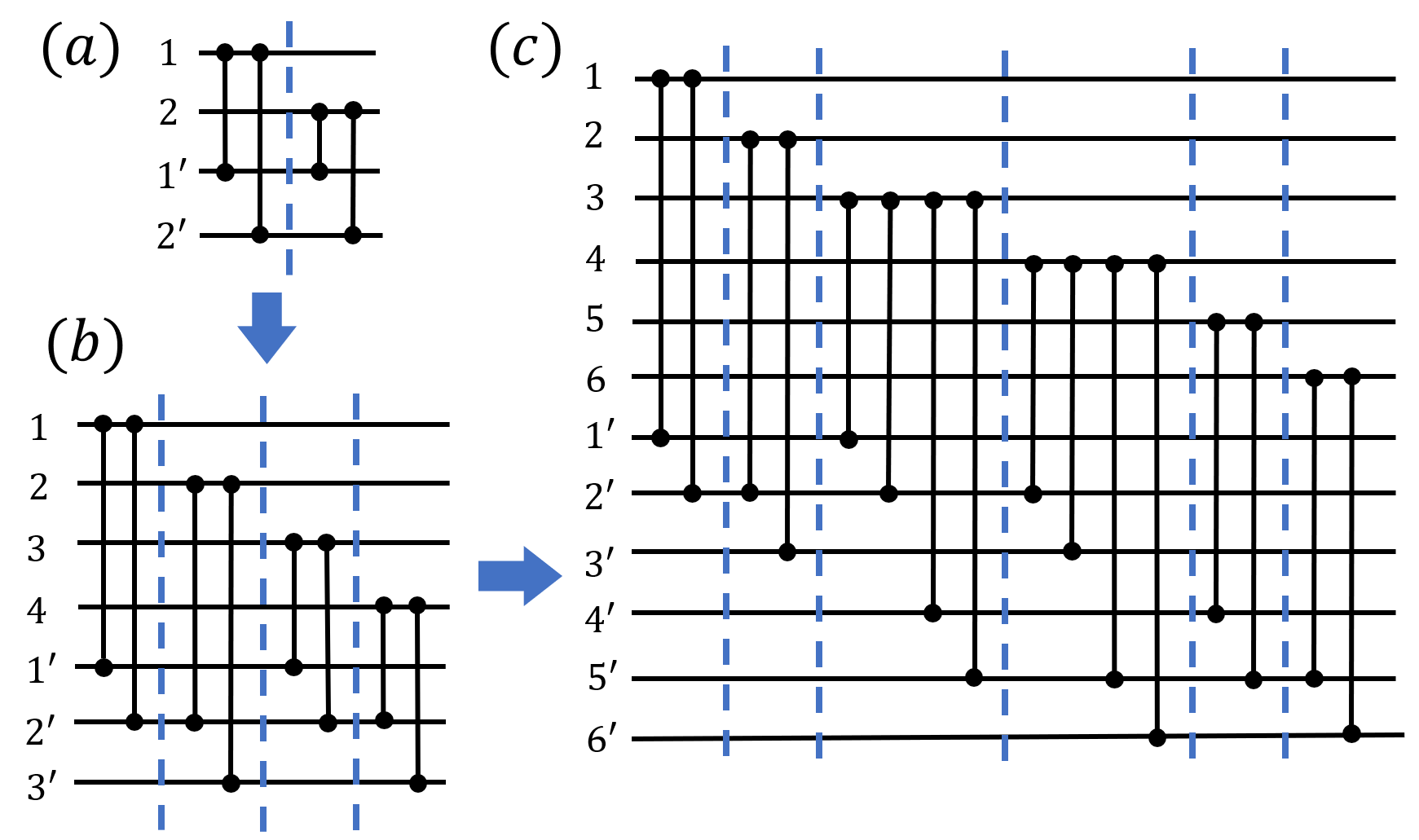}
\caption{The circuits for generating the cluster states in Fig. \ref{Fig:Fact.Eg.3} (a), (b), and (c). Each one of them is actually the quantum circuit version of the cluster state with the corresponding label. The quantum gates in the circuits are only the CZ gates.}\label{Fig:Fact.Eg.3-1}
\end{figure}

\begin{figure}
\centering
\includegraphics[width=3.4in]{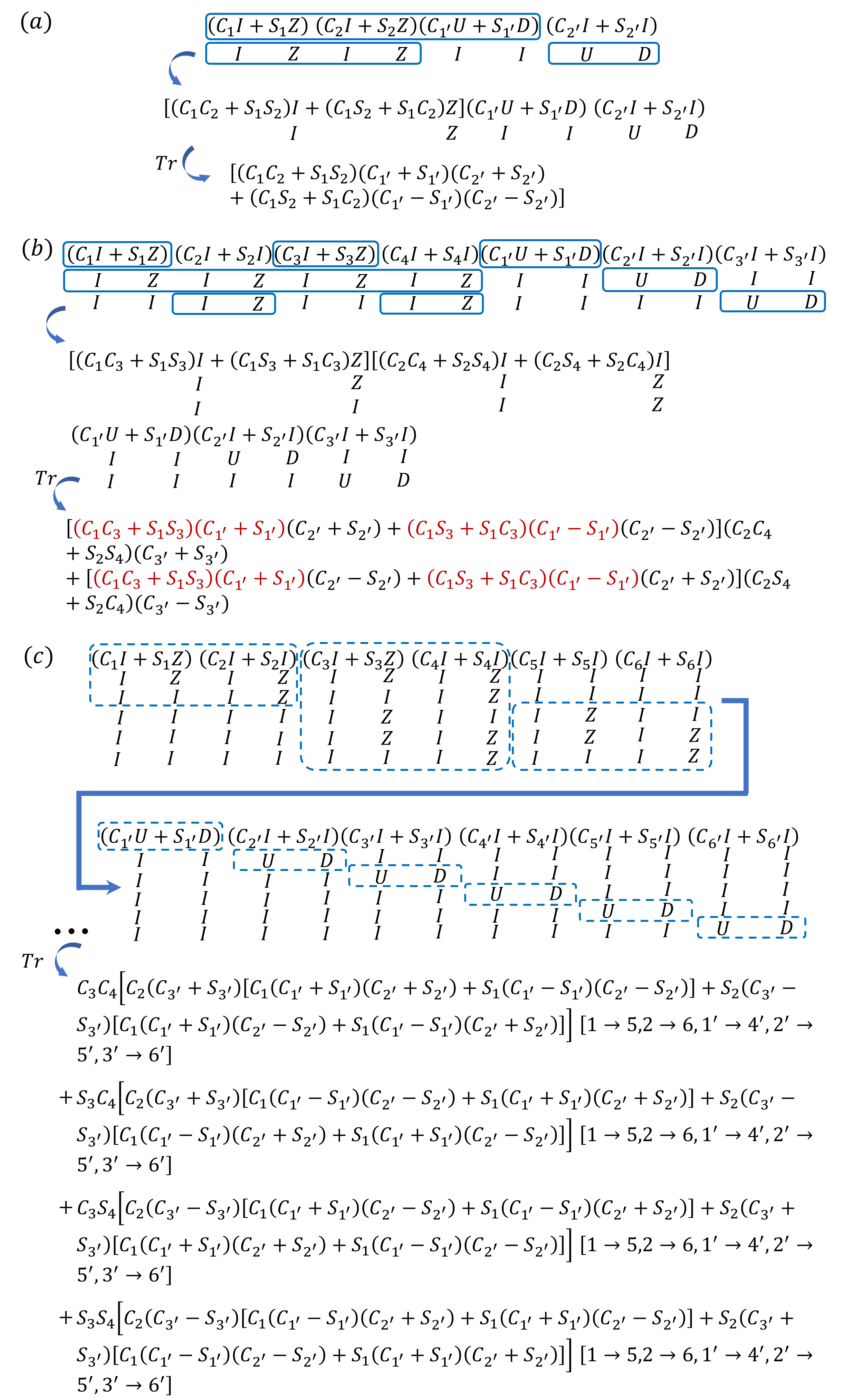}
\caption{The polynomials for computing the cluster states in Fig. \ref{Fig:Fact.Eg.3} (a), (b), and (c), followed by brief instructions on how to obtain the recursive relations. Each one of them is for the cluster state with the corresponding label. The blue boxes (both dashed and solid ones) are used for indicating the ``block'' structure of the tensors. The red terms in (b) are the ones that need special notice when observing the patterns (like those colored ones in Fig. \ref{Fig:Refac-3} and \ref{Fig:Refac-4}). In (c), only the final expression is given. The terms in the square brackets are in the same form as those in the square brackets before them, respectively, but with different subscripts. The small black arrows in the square brackets indicates the transformation of the subscripts of the terms inside square brackets from those in the front brackets.}\label{Fig:Fact.Eg.3-2}
\end{figure}

The value of each point in the curves is an average over 25 trials, and its standard deviation is also marked out in the plot. In one trial, a random projection direction is chosen, and the particular strategy to choose it is the same as that applied in the second example. For one cluster state among those in Fig. \ref{Fig:Fact.Eg.3} (a), (b), and (c), the projection direction is the same for the three algorithms in one trial. The blue curve is obtained by generating complex state vectors of the three cluster states, and the circuits for the generation are shown in Fig. \ref{Fig:Fact.Eg.3-1}. The actual actions are simple. One can first generate the tensor product of the state vector $|+\rangle$, and then apply the networks of the CZ-gates, as shown in the three panels of Fig. \ref{Fig:Fact.Eg.3-1}, to them. The circuit for generating the corresponding cluster state in Fig. \ref{Fig:Fact.Eg.3} is labeled the same. The green and pink curves are obtained by the factorized polynomial, which are shown in Fig. \ref{Fig:Fact.Eg.3-2}. In particular, the green curve is obtained by computing the trace of the tensor form of the polynomial. Those forms are given in the top rows of panels (a), (b), and (c), respectively, which can be viewed as some sort of ansatz to the projections. The pink curve is obtained by using the recursive relations derived from the polynomials, which are given in the lower rows of the three panels. In addition, the polynomial and the derivation for computing the corresponding cluster state in Fig. \ref{Fig:Fact.Eg.3} are labeled with the same letters.

From the curves in Fig. \ref{Fig:Fact.Eg.3}, a clear reduction in running time can be observed. The data points in the figure can be given precisely. For the blue curve, the coordinates of the three points are $(4,-8.0880)$, $(7,-6.1594)$, and $(12,3.4323)$, and their standard deviations are $1.3441$, $0.3257$, and $0.0043$, respectively. As noted above, the vertical coordinates are the natural logarithm of the running time (in seconds), so one needs to take the exponential of the data to obtain the running time. For example, the actual running time of the 4-qubit case (the cluster state in Fig. \ref{Fig:Fact.Eg.3}(a)) is $e^{-8.0880}=3.0720\times10^{-4}$ seconds, and the rest are similar. The three points of the green curve are $(4,-8.3574)$, $(7,-7.0639)$, and $(12,-5.6397)$ with the standard deviations $0.8786$, $0.1195$, and $0.8209$ correspondingly. The three points of the pink curve are $(4,-13.8708)$, $(7,-13.0013)$, and $(12,-11.5978)$ with the standard deviations $0.5125$, $0.0973$, and $0.1284$ correspondingly. For the blue curve, one may easily find that its increase rate with qubit number is faster than proportional relations, indicating that the running time of this algorithm will grow exponentially with the qubit number. For the green and pink curves, they look like straight lines, but are actually bent down. This can be seen more clearly by fitting lines with the first two points of the curves and locating the third point in reference to the fitted line. For the green curve, the third point is below the fitted line based on the first two points (the fitted slope is $0.4312$). The vertical coordinate difference between the third point and the point on the fitted line with the same horizontal coordinate is $-0.7318$. For the pink curve, the third point is also below the corresponding point on the fitted line (the fitted slope is $0.2898$) in the vertical direction with a difference $-0.0458$. These indicate that the running time of the algorithms based on the factorized polynomials have a clear advance over the direct computational strategy based on exponentially large complex vectors for evaluating the local projections on 2DSLCSs. Of course, the author realizes that there are many optimized algorithms for such computational problems, which might also be quite efficient. In addition, the scalings shown in Fig. \ref{Fig:Fact.Eg.3} are not sufficient to verify the conclusions in Section \ref{sec:CS} due to the lack of data. Although the slopes of the fitted line are quite small, the running time scale indicated by the green and pink curves could also be $e^{N^{\gamma}}$ (with the qubit number $N$ and the real number $\gamma<1$) \cite{Footnote2}. Of course, the green and pink curves are expected to be $\ln{N}$, leading to the polynomial scaling of the running time in terms of the qubit number. This is a bit difficult to identify by the data in Fig. \ref{Fig:Fact.Eg.3}. To this extent, we think the above results are worth sharing.    

\section{Potential Connections and Usages}
In this section, several thoughts about the method in this note is presented, including potential connections to other methods and applications. 

1) Again, we realize that the method in this note is very similar to the tensor network method for quantum computing \cite{Berezutskii2025}. Especially, a matrix product state (MPS) in the tool box is defined by \cite{Orus2014}
\begin{equation}
    |\psi_{MPS}\rangle=\sum_{j_1,j_2,\ldots,j_N}^1\Theta_{j_1,j_2,\ldots,j_N}|j_1,\dots,j_N\rangle, 
\end{equation}
where 
\begin{equation}
    \Theta_{j_1,j_2,\ldots,j_N}=\sum_{\alpha,\beta,\ldots}\mathcal{A}_{\alpha,j_1,j_2}\mathcal{B}_{\beta,j_3,j_4}\ldots. 
\end{equation}
In other words, the basis amplitudes of an MPS can be decomposed into the product of finite matrices. This allows an efficient computation of the amplitudes using parallel classical algorithms. The method in this note seems to belong to a subcategory of the MPS method, despite that the two methods come from different backgrounds. Besides, for a general MPS, one does not necessarily obtain a factorized polynomial as shown in this note. It seems that the method in this note is more efficient, is it?

2) It is known that a 2DSLCS is one kind of universal quantum computing resource, meaning that any type of quantum algorithm can be implemented using the state. To the best of our knowledge, only local projectors are required for performing universal quantum computing. If this is precise, a polynomial strategy for computing the projection results of 2DSLCS could be viewed as a polynomial ``shortcut'' to perform all quantum algorithms. This might cause a collapse in some place of the hierarchical structure of computational complexity. Again, we do not major in mathematics or theoretical computation. So, we are not sure about whether there is a mistake in the above point, is there? In Appendix \ref{sec:append}, a fundamental strategy is shown that can transfer circuits of control-phase gates (or normally CPhase gates) to cluster states. The cluster states for more complicated quantum circuits can also be constructed using the tools in Appendix \ref{sec:append}.  

3) A famous example of the quantum advance is the efficiency of running the Shor's algorithm \cite{Shor1999} for integer factorization. One typical character of quantum states involved in the Shor's algorithm is that, for the output state of the quantum Fourier transform (QFT) subroutine, the measurements on the state would give the period of the modular exponential function with the highest chance. This sounds like a one-shot action, but it is actually not. The reason is that, when sampling a quantum state, the distribution (or equivalently, the basis amplitudes) cannot be obtained by a single sampling outcome. In the best situation of the method in this note, a polynomial strategy might be given for obtaining each one of the basis amplitudes of the QFT output state in the Shor's algorithm---one basis amplitude corresponds to one setup for the projection direction. However, there would be $2^N$ basis states for the $N$-qubit case, so that it requires exponential trials to obtain all basis amplitudes. In the author's opinion, given a polynomial strategy for computing each basis amplitude, the exact computations of all the basis amplitudes can be replaced by a classical random sampling process. In particular, one can randomly pick out a set of directions at first, each of which corresponds to a basis state. Then, calculate the projection results of them using the polynomial strategy. If the set of randomly selected directions is sufficiently large, one can also estimate the period of the modular exponential function with a bounded error. This is like randomly sampling a classical algorithm or circuits. The analyses on similar sampling for bosonic circuits are given in Ref. \cite{Oh2024-2}. In all, we think that the Shor's algorithm could also be efficiently performed using the strategy in this note, could it? 

4) Going a bit further, it is well known that the quantum advantage in query complexity represents a significant revolution in algorithm design. A prime example is Simon's algorithm \cite{Simon1997}, which demonstrates an exponential speedup compared to its classical counterparts and serves as the inspiration of Shor's algorithm \cite{Wiki-1}. Crucially, it can be rigorously proven that while Simon's algorithm solves the problem with $O(n)$ queries (the query number is asymptotically bounded above by the size of the problem $n$, up to a constant factor), any classical algorithm necessarily requires $\Omega(2^{n/2})$ queries (the query number is asymptotically bounded below by $2^{n/2}$) to achieve the same task. This represents an exponential separation that holds unconditionally, without relying on complexity-theoretic assumptions like $P\ne BQP$. The above advantage fundamentally stems from the unique capabilities of quantum mechanics, which allow for the manipulation of probability amplitudes with both positive and negative phases. Specifically, the algorithm ensures that the measurement results orthogonal to the target vector $\vec{s}$ (which indicates the hidden period) acquire destructive interference through precisely engineered amplitude cancellation, resulting in a strictly zero probability of observing such results. The critical insight lies in verifying whether the amplitude magnitude of specific expansion terms falls below a threshold with a probability exceeding $2/3$. Remarkably, this verification process circumvents the experimental challenges associated with preparing full quantum superposition states. As a result, the demonstration of quantum advantage is possible even in classical wave systems that support phase cancellation phenomena, such as destructive electromagnetic wave interference. 

5) We notice that machine learning can also be performed using the strategy in the note. As indicated in Ref. \cite{Sun2024}, the learning of a data set can be performed using a 2DSLCS. In such a formulism, the input data is encoded by the qubits of the 2DSLCS. Compared with the total state, the number of qubits that carry input data is very small. The outputs are defined by the projection on the 2DSLCS, and the training parameters are given by the angles of the projection direction. Because the measurements on quantum states are usually independent of the setup for generating the states,  one benefit of the learning scheme based on 2DSLCSs is that the setup for the total state does not need to tune much during the learning process once the structure is settled. This might suppress the systemic error induced by the dynamical controls. Finally, by optimizing the projection direction, the desired output can be finally obtained. 
In Ref. \cite{Sun2024}, it is shown that an improvement in training efficiency can be observed by comparing the performance of the learning based on 2DSLCSs with those previous methods. This is at least valid for some simple datasets. Hence, the polynomial strategy in this note could provide a classical but efficient algorithm as a subroutine for the learning scheme discussed in Ref. \cite{Sun2024}.

\begin{figure}
\centering
\includegraphics[width=3.4in]{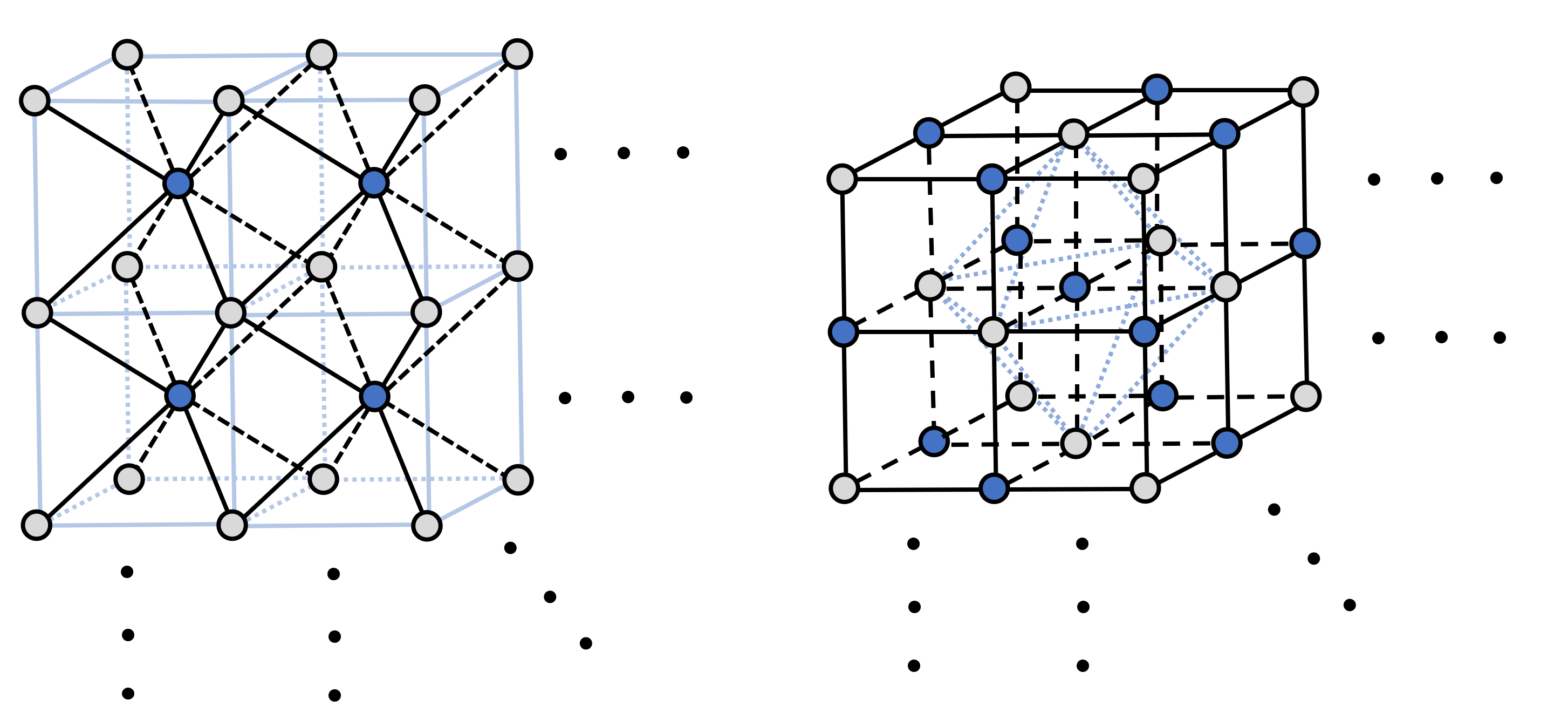}
\caption{Two kinds of 3-D cubic lattice cluster states. The unit cluster state of the left (right) one is a 9-qubit (7-qubit) GHZ state.  }\label{Fig:3Dcase}
\end{figure}

6) Despite being a bit far from the topic, we would like to mention the 3-D Ising model \cite{Viswanathan2022}. In fact, the idea of MBQC is related to the investigations on 2-D Ising models. In an Ising model, the interactions among spins are Z-Z couplings. Hence, the evolution operator of a pair of spins for a given time is also known as ZZ rotation gates in quantum computing (and that is why they are also called Ising ZZ gates). For 2DSLCSs, the interactions that generate them from the tensor product states are CZ gates, which are close to the ZZ gates \cite{Formula-1}. So, perhaps, the trick in this note would be helpful for investigating Ising models as well. Indeed, calculating the projections of the cluster states is different from calculating the eigenvalues or partition function of Ising Hamiltonian. However, the idea of renormalization asartz to the 2-D Ising models is like the attempts for the recursive relations in the note. They both focus on the symmetry of the theory in terms of the system size. Meanwhile, the method in the note could be extended to the 3-D case as well. Two kinds of 3-D cluster state are shown in Fig. \ref{Fig:3Dcase}. They also share some similarities with the 2DSLCSs discussed in the note: 1) all the connections between qubits are only CZ gates, and 2) the number of qubits that directly connect to one single qubit is limited (the number is 4 for 2DSLCSs, and the number could be 6 or 8 for 3-D cases). Therefore, one could also establish a factorized polynomial for the local projections on those 3-D cluster states, or precisely the 3-D cubic lattice cluster states. In that case, instead of a cross-shape cluster state, the unit would be a star-shape cluster state. It has one qubit in the center and 6 or 8 disconnected qubits linked to the center individually with CZ operations. Like the cross-shape cluster state, such a state is also a GHZ state. Therefore, the basic product of factors would be composed of 7 or 9 binomials with order one, instead of the 5-binomial case for 2DSLCSs. Hence, we guess that such a factorized polynomial for 3-D cubic lattice cluster states might be helpful for investigating the exact solutions to 3-D Ising models.

7) Finally, we would like to mention another interesting but not-so-well-known theorem called ``freshman'' theorem (or ``a freshman's dream''). The story comes from a common mistake for freshmen in computing the expression $(x+y)^n$. Certainly, if one distributes all the terms across the brackets, the result is $\sum_{l=0}^nC_l^nx^ly^{n-l}$, rather than $x^n+y^n$. However, one could ask whether there is a case where $(x+y)^n=x^n+y^n$. Actually, if $x$ and $y$ belong to the ring of integers modulo $p$, where $p$ is a prime number, the equation will hold. Formally, it is expressed by
\begin{equation}\label{eq:Freshman}
    (x+y)^p=x^p+y^p,~\forall~x,y\in\mathbb{Z}_p.
\end{equation}
In fact, many quantum phenomena, such as quantum entanglement, have a deep connection with the formula. For example, according to Greenberger-Horne-Zeilinger theorem, the correlation of the GHZ state cannot be given by simply multiplying the local measurement results of all the particles. Compared with Eq. (\ref{eq:Freshman}), if one thinks of $(x+y)^p$ as some sort of multiplication of ``local measurements'' and $x^p+y^p$ as some sort of ``correlations'', the GHZ theorem could be understood in a different way. Of course, this is not a strict statement, because $x$ and $y$ in Eq. (\ref{eq:Freshman}) are integers, while the measurement results of the GHZ state are real numbers. We do not know whether $\mathbb{Z}_p$ is the only set that makes the equation hold, probability not. However, the point is that the boundary of the quantum and classical world is very hard to address. As shown by the above vague relation, one might find a classical system that can be described by $\mathbb{Z}_p$, right? In a joking way, a classical computer could be such a system, for the fact that it can generate such a set of $\mathbb{Z}_p$. Thus, one perhaps could not claim a system to be quantum by merely one theorem. The interesting thing is that, in the microscopic world, there are some phenomena that cannot be described by the commonly used math ---the algebraic structure of real numbers. It might not give the conclusion that the phenomena described by different math, such as an enhanced version of the ring for the``freshman's dream'', uniquely happen in the microscopic world. In complicated classical systems, such as fluid or irregular scatters of electromagnetic waves, there should be measurement results that could be described by the not-commonly-used mathematical structure. In our personal view, we think that the lack of tools for the observation of the microscopic world gives birth to many wild imaginations about the reality of that world. It is definitely good to imagine it in various ways, but it might not be that good to turn those imaginations into beliefs of reality without any doubt or reservation.  

\section{Perspectives}
This part concludes the note. In brief, we present a primary note on one kind of polynomial classical algorithms for evaluating the local projections on one kind of universal quantum computing source---2-D square lattice cluster states. Furthermore, the step number for performing the algorithm is likely to be proportional to the number of qubits in the cluster states. Considering that nearly all the quantum algorithms can be performed by setting local projectors for a 2-D square lattice cluster state---if we understand it correctly, the trick in the note might bring new understanding of the computational architecture illustrated by the quantum computing strategy.  

Of course, there might be flaws in this note. Even if it is correct, similar tricks might already have been found in other researches. After all, there are so many talents out there, and our knowledge is quite limited. So, the motivation for writing this note is just the intention to share and communicate. The math in this note is not complicated, so the formulas and results are easy to check. If the results of the note turn out to mean something, we hope that they could provide a broad playground for researchers who like them. If the results of the note turn out to mean nothing, just throw them behind, like what we do to many publications. 

A little re-emphasizes on the point in the preface of the note: In a more broad sense, it should not be surprising that a classical system has properties similar to those of a quantum system. No matter how large the gap between the two systems is, they are both described by mathematics found by us humans. The two pillars of quantum mechanics theory, linear algebra and statistical theory, have been developed and applied for hundreds of years. In our opinion, the true surprising thing should be that the physical law of the microscopic world can still be described by the mathematical tool found for hundreds of years. It does not prove that what we found hundreds of years ago is the truth of the world. It might just prove that some of the investigations in the past century are really in short of imaginations. Unfortunately, the phenomenon that cannot be described by the math tool we found would also not be strictly understandable for us. The basic requirements for being ``strictly understandable for us'' should be (1) that we can observe the phenomenon again and again under a specific condition, and (2) that we could potentially generalize the precise description of the phenomenon. In recent years, multilayer machine learning models have become a new fashion in almost all fields of science. Maybe, it represents a temporary victory of the recursive-relation-based computing strategies that can connect the finite to the infinite---after all, there are many theories derived from the finite-sized problem failing to solve the infinite-sized problem ultimately. However, whether it could broaden the boundaries of our knowledge on physical world is still unknown.

Returning to computational problems, the results in this note are also based on the fundamental math tool of linear algebra. This is also one of the constraints of such computations. So, what kind of constraints would you like to choose? Then, how many steps would you like to take to perform the computations under the constrains? \cite{Wolfram2021} Cheers to dancing in chains, right?

Ending with a famous quote from the book {\it Tractatus Logico-Philosophicus} written by Ludwig J. J. Wittgenstein:
\begin{displayquote}
Die Grenzen meiner Sprache bedeuten die Grenzen meiner Welt.
\end{displayquote}
It means ``The limits of my language mean the limits of my world'' in English. In fact, the limits of math might do the same to the limits of our knowledge of the physical world.
\begin{figure}
\centering
\includegraphics[width=3.4in]{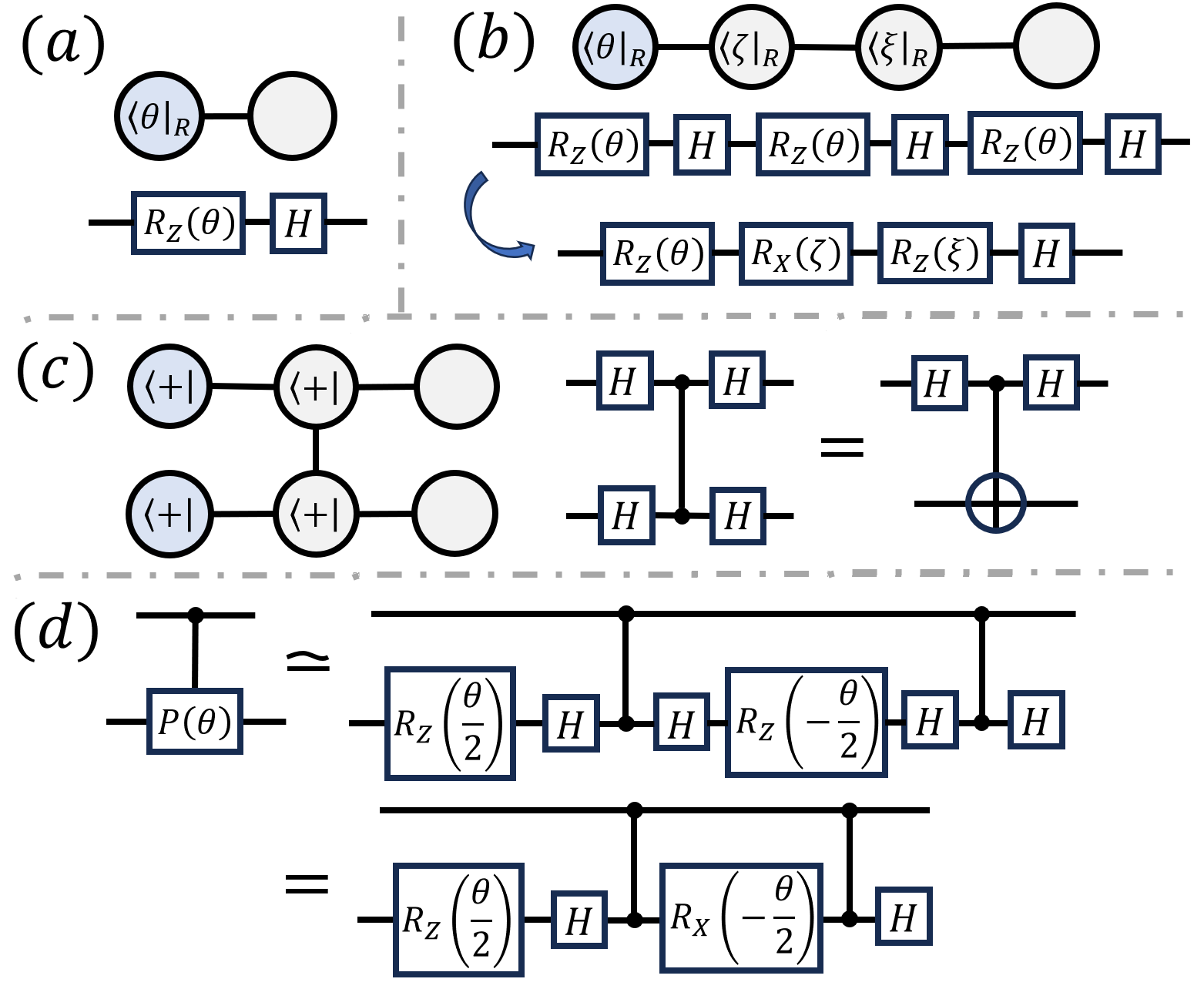}
\caption{The mapping relations of MBQC and a universal quantum gate set, and a decomposition of CPhase gate. In (a), (b), and (c), the gray circles denote the qubit state $|+\rangle$, and blue circles denote the input states $|\psi\rangle$ (in (a) and (b)) or $|\psi\rangle\otimes|\varphi\rangle$ (in (c)). The projectors of the qubits are also marked on the circles. Those ones with no projectors marked sever as the outputs. Hence, (a), (b), and (c) presents the MBQC scheme of a Pauli-Z rotation (followed by a Hadamard gate), an arbitrary rotation (followed by a Hadamard gate), and a CNOT gate (with its controls in the Pauli-X Basis), respectively. The quantum circuits are given near the correponding cluster states. In (d), a decomposition of CPhase gate is presented. The notation $\simeq$ indicates that the circuits on the both sides have a global phase difference, which is not relevant.}\label{Fig:Toolbox1}
\end{figure}
\appendix
\section{One set of tools for transferring a quantum circuit into a 2D cluster state}\label{sec:append}
\begin{figure}
\centering
\includegraphics[width=3.4in]{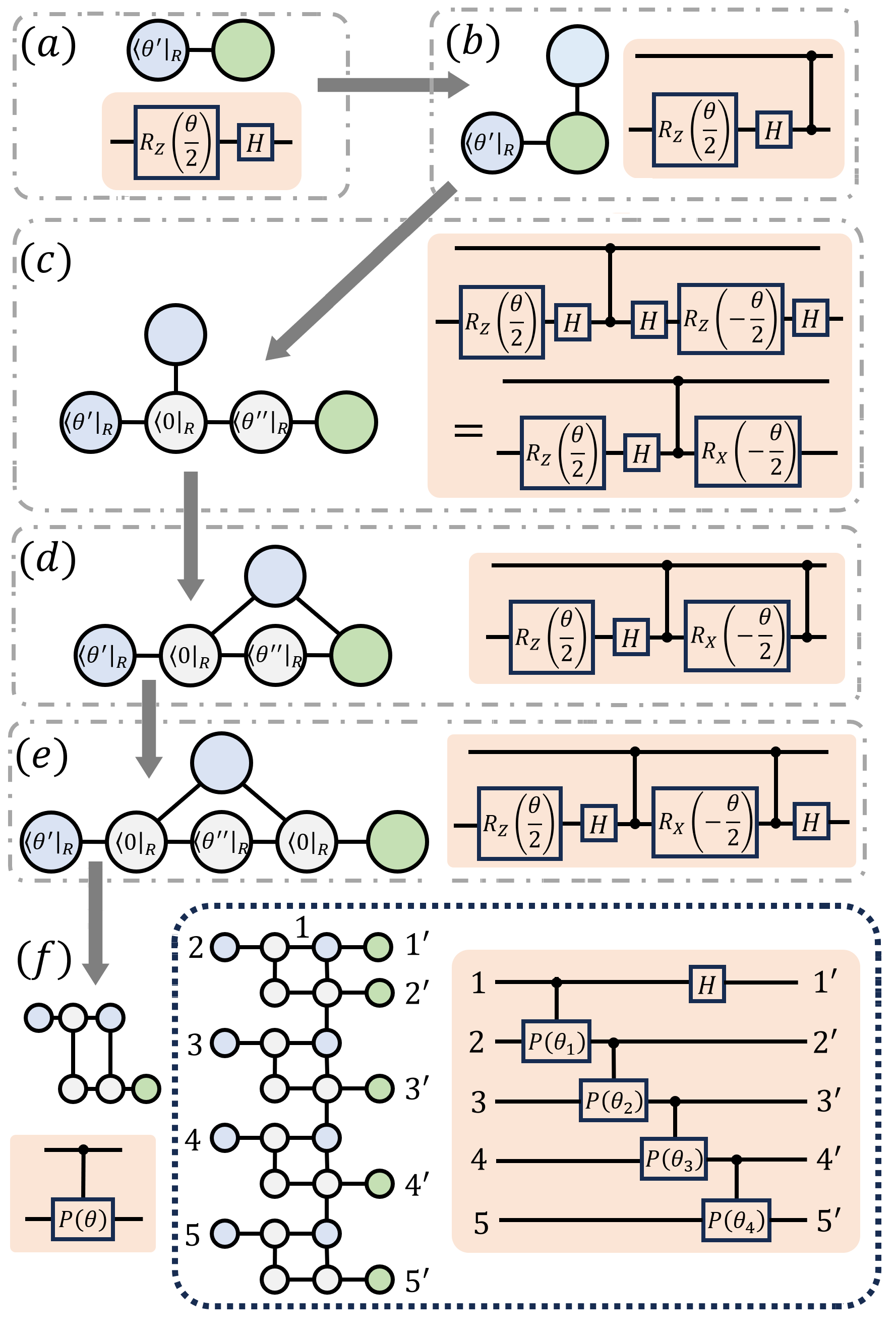}
\caption{A step-by-step procedure for obtaining the cluster state and the projectors for performing a CPhase gate, and the composition of the cluster states. (a) to (e) illustrate the procedure. The circles without a projector serve as the outputs. The corresponding quantum circuits are presented by the side of the cluster states. (f) illustrates the composition of the cluster states for performing the circuit composed of CPhase gates.
}\label{Fig:Toolbox2}
\end{figure}
There have been plenty of technical papers on how to decompose a circuit, as well as how to map a quantum circuit to a cluster state. In the appendix, we present a brief illustration of the approaches for them, as shown in Fig. \ref{Fig:Toolbox1}. Fig. \ref{Fig:Toolbox1} (a) to (c) are the mapping relations between several fundamental quantum gates and cluster states \cite{Campbell2009}. Fig. \ref{Fig:Toolbox1} (a) shows a two-qubit cluster state at the top, where the blue circle denotes the input qubit $|\psi\rangle=\alpha|0\rangle+\beta|1\rangle$, and the gray circle represent $|+\rangle$. $\langle\theta|_R$ is the projector on the ``blue'' qubit, defined by $\langle\theta|_R=\langle0|He^{-i\theta Z}$. $H$ is the 2-by-2 Hadamard operator, and $Z$ is the Pauli-Z operator. Then, one has
\begin{equation}
    \begin{split}
        &(\langle\theta|_R\otimes I)CZ(|\psi\rangle\otimes|+\rangle)\\
        =&(\langle0|\otimes I)(He^{-i\theta Z}\otimes I)CZ(|\psi\rangle|+\rangle)\\
        =&(\langle0|\otimes I)(H\otimes I)(e^{i\theta Z}\otimes I)CZ|\psi\rangle|+\rangle\\
        =&(\langle0|\otimes I)|(H\otimes I)CZ(e^{i\theta Z}|\psi\rangle)|+\rangle\\
        =&(\langle0|\otimes I)|\frac{1}{\sqrt{2}}(|0\rangle+|1\rangle X)(\alpha|+\rangle+\beta|-\rangle)\\
        =&He^{-i\theta Z}|\psi\rangle,
    \end{split}
\end{equation}
where $X$ is the Pauli-X operator. Therefore, by projecting the ``blue'' qubit on $\langle\theta|_R$, the ``gray'' qubit will be turned to $He^{-i\theta Z}|\psi\rangle$, which is equivalent to the circuit below the cluster state in Fig. \ref{Fig:Toolbox1} (a). Here, $R_Z(\theta)=e^{-i\theta Z}$ (sometimes, $R_Z(\theta)$ is defined by $e^{-i\theta Z/2}$). Such a simple scheme is also called single-qubit teleportation protocol. Afterwards, an arbitrary single qubit rotation can be realized by projecting the first three qubits of a four-qubit-line-shape cluster state by $\langle\theta|_R\langle\zeta|_R\langle\xi|_R$, as shown at the top of Fig. \ref{Fig:Toolbox1} (b). The ``blue'' qubit is also the input state $|\psi\rangle$, and the rest qubits are $|+\rangle$s. After the projection, the fourth qubit is turned to $HR_Z(\xi)R_X(\zeta)R_Z(\theta)|\psi\rangle$, where $R_X(\zeta)=e^{-i\zeta X}$. The circuit equivalent to the projection is given below the cluster state in Fig. \ref{Fig:Toolbox1} (b). Furthermore, because of the property of a CZ gate, a CNOT gate can be directly realized by the ``I-shape'' cluster state in Fig. \ref{Fig:Toolbox1} (c). The two ``blue'' qubits represent the input state $|\psi\rangle\otimes|\varphi\rangle$, and the remaining qubits are also $|+\rangle$s. After projecting the ``I-shape'' cluster state on $(\langle0|H)^{\otimes N}=\langle+|^{\otimes N}$, the ``gray'' qubits in Fig. \ref{Fig:Toolbox1} (c) without being projected are turned to $CZ|\psi\rangle|\varphi\rangle$. The circuit equivalent to the projection is given on the right side of the cluster state in Fig. \ref{Fig:Toolbox1} (c). Collecting the relations Fig. \ref{Fig:Toolbox1} (a), (b), and (c), all kinds of quantum circuit can be realized by projecting a cluster state on the tensor product of $\langle\theta|_R$s with suitable $\theta$s (notice that $\langle0|_R=\langle0|He^{-i0 Z}=\langle+|$), based on the concept of universal quantum gate set. Especially, the projection on $\langle+|$ is equivalent to performing a single Hadamard gate on the qubits connected to the projected one, and then removing the projected one (with the lines). Thus, one can remove (or add) the lines and qubits from a cluster state according to the rules, in case that it is required by the realization of a circuit. The same goes for the 2DSLCS in this note. A detailed instruction of the above tools can be found in Ref. \cite{Campbell2009}.

Next, we present an example of how to obtain the corresponding cluster states and projectors for a given quantum circuit. The example here is very simple, and the key ingredient is the decomposition given by Fig. \ref{Fig:Toolbox1} (d) \cite{Crooks2024}. The cluster state for performing a CPhase gate can be obtained by the decomposition. A step-by-step procedure is shown in Fig. \ref{Fig:Toolbox2} (a) to (e). The gray arrows indicate the transformation of the state. Similarly to Fig. \ref{Fig:Toolbox1}, the blue and gray circles also denote input qubits and $|+\rangle$, respectively. The green circles denote the qubits as the output. Because the control qubit in the CPhase circuit is not operated by any gate, the blue circles with no projectors also serve as the output (which should be colored green, but, never mind). The particular parameters satisfy $\theta'=\theta/2$ and $\theta''=-\theta/2$. The quantum circuits of the corresponding measurement-based schemes are given near them. Notice that the final cluster state for a CPhase gate, given in Fig. \ref{Fig:Toolbox2} (e), is a square-shape cluster state with two individual qubits connected its two diagonal qubits respectively, as shown in Fig. \ref{Fig:Toolbox2} (f). Therefore, a five-qubit circuit composed of four CPhase gates acting on adjacent qubits (with an extra Hadamard gate) can be mapped to a connection of the cluster state for a CPhase gate, as shown in the dashed box of (f). Order numbers 1 to 5 are used for input qubits, and $1'$ to $5'$ for output qubits \cite{Footnote3}. As we have mentioned above, projecting a qubit of a cluster state by $\langle+|$ is equivalent to performing a Hadamard operator on the qubits connected to it, and then removing it together with the connections. Therefore, one can make the cluster state in the dashed box of (f) from a 2DLSCS (like carving a pattern on a stone). Actually, the circuit in the dashed box of (f) can be thought of as a basic component of the quantum Fourier transform circuit (after reordering the gates properly).


\begin{thebibliography}{99}
\bibitem{Footnote1}
Actually, many of the Feynman's ideas have shown the spirit of question. An example is the Feynman's trick on integration. For the integral like $\int_1^{\infty}x^5e^xdx$, one can define a function $I(a)=\int_1^{\infty}e^{-ax}dx=-e^a/a$. Then, the integral can be evaluated by computing the fifth order derivative of $I(a)$ relative to $a$, which is much easier than directly computing the original integral. The subspace that contains the functions like $I(a)$ could be thought as a ``good choice of space'' for computing the integrals like $\int_1^{\infty}x^5e^xdx$.
\bibitem{Footnote-1}
The factorization of the polynomial seems not to be valid on the complex number field either, i.e., $a,b,c,d\in\mathbb{C}$. The author does not check the case strictly. It can be seen from the solvability of the equation set obtained by expanding the assumed factorized form and letting the coefficients of the corresponding terms equal to each other.
\bibitem{Footnote-Insert}
A problem of recursive algorithms is that the error, if any, in one step will accumulate easily through out the whole calculation. A typical example is the interpolation and approximation (of a polynomial) based on Chebyshev polynomials. It is known that there are a series of methods for suppressing the error in recursive algorithms, such as controlling the weights of recursive steps. They will also be helpful for specific coding based on the strategy in this note.
\bibitem{Footnote2}
If the green and pink curves could be fitted by $N^{\gamma}$, the scaling of the running time would be $e^{N^{\gamma}}$. However, a curve should have tangent lines with slopes bigger than 1 if it could be fitted by $N^{\gamma}$. Albeit the slopes of the fitted lines here are far less than 1, it is still hard to draw a strict conclusion.
\bibitem{Footnote3}
Please do not mix them with the notations in Section \ref{subsec:CSCS} and \ref{subsec:2DCSCS}, and Section \ref{sec:egs}. They are differently defined in their sections, respectively.
\end{thebibliography}
\end{document}